\documentclass{jfm}

\usepackage{graphicx}
\graphicspath{{figs/}}
\usepackage{newtxtext}
\usepackage{newtxmath}
\usepackage{amsmath, amssymb}
\usepackage{bm}
\usepackage{natbib}
\usepackage{hyperref}
\usepackage{subcaption}
\hypersetup{
    colorlinks = true,
    urlcolor   = blue,
    citecolor  = black,
}

\newcommand{\RomanNumeralCaps}[1]
\linenumbers

\newcommand{\C}{$^\circ$C}
\newcommand{\id}{\mathrm{d}}

\title{Wave erosion of ice cliffs: melt rate due to reflection of non-breaking surface waves}

\author{Anya Wolterman\aff{1},
  Till J. W. Wagner\aff{2},
  Lucas K. Zoet\aff{3}
 \and Nimish Pujara\aff{1,4}  \corresp{\email{npujara@wisc.edu}}}

\affiliation{
\aff{1}Department of Civil and Environmental Engineering, University of Wisconsin-Madison, Madison, WI 53706, USA
\aff{2}Department of Atmospheric and Oceanic Sciences, University of Wisconsin-Madison, Madison, WI 53706 USA
\aff{3}Department of Geoscience, University of Wisconsin-Madison, Madison, WI 53706 USA
\aff{4}Department of Civil Engineering, The University of British Columbia, 2002-6250 Applied Science Lane, Vancouver, BC V6T 1Z4, Canada
}

\begin{document}
\maketitle

\begin{abstract}
Wave erosion of ice cliffs is one of the main mechanisms for waterline ice melt for icebergs, glacier fronts, and ice-shelf fronts. Despite its importance, this process is fundamentally not well understood or extensively tested in controlled experiments, and therefore coarsely parameterized in geophysical and climate models. In this study, we examine the melting of a vertical ice wall caused by surface waves using both theory and laboratory experiments, with an emphasis on the flow-induced heat transport in the theory and measurements of the melt rate profile under different wave conditions in the experiments. In both the theory and the experiments, we find that there is enhanced melting near the surface that decays with depth. Via an analysis the oscillatory boundary layer flow, we find that an approximate, leading-order, wave-averaged balance of heat transport is given by the vertical advection due to an Eulerian boundary layer streaming current and horizontal diffusion. By solving for this balance and obtaining the wave-averaged temperature field, we find an explicit expression for the wave-induced melt rate. Experimental data show a good match to this expression, especially for larger wave amplitudes and colder water temperatures, though we find that the ambient melt can be a significant contributor to the waterline melt rate. 
\end{abstract}



\section{Introduction}
\label{sec:introduction}

The ice mass within glaciers, ice sheets, and icebergs constitutes an important, but delicate component of the global energy balance. Small changes in the ice mass fluxes can have large impacts far beyond the high-latitude regions where this ice predominantly resides. In ice--ocean interactions, the ice melt that occurs at this interface is typically controlled by the fluid motion in the ocean. \citet{Du24} provide a useful perspective on the recently developed laboratory, computational, and theoretical tools being used to study this coupled process from a fundamental fluid mechanical perspective, and how the results can be connected with processes that occur at larger, geophysical scales relevant for climate models.

Of the myriad processes occuring at the ice--ocean interface, we focus here on how surface gravity waves can erode the sidewalls of icebergs, the faces of marine and freshwater-terminating glaciers, and the fronts of ice shelves. The efficacy with which wave-induced melt occurs depends on numerous environmental factors that vary widely between different settings: the amplitude and period of the incoming waves; whether the ice face is exposed to standing or breaking waves; the stratification and temperature profiles of the water; whether the water surface is covered by sea ice or an ice mélange; and the material integrity and geometric configuration of the ice face itself. However, how each of these impact the erosion rate remains poorly understood. Challenges when studying this process include: in situ observations of ice cliffs are costly, cumbersome, and risky; laboratory investigations require careful control of ice, water, and wave conditions, ideally in cold ambient temperatures; and theoretical progress is made difficult by the complex interactions at the ice--water interface, with confounding roles of temperature, salinity, and wave action. 

From observations, we know that wave-driven erosion is typically maximized for icebergs floating in the open ocean, particularly once they have drifted to lower latitudes. At the other end of the spectrum, wave action is largely suppressed in mélange-choked fjords. However, conditions can quickly change, for example when loss of sea ice exposes the glacier or ice shelf front to open water, allowing wave erosion to become an important factor in determining calving rates. Wave-induced erosion has historically been partitioned into three main processes \citep{Savage:2001hz}: (i) wave-induced melt resulting in incision near the waterline; (ii) calving of the above-water section of the ice cliff that overhangs the waterline incision; (iii) submerged or full-depth calving due to buoyant forces on underwater ice benches (or ``feet''), produced by the collapse of the above-water cliff, that can protrude on the order of 100 m beyond the overhang. Both (ii) and (iii) respond directly to the rate of the wave-induced melt (i) and hence should scale with it, making this particular melt process a potential key driver of ice loss for icebergs, glaciers, and ice shelves.

\citet{White:1980up} is the prevailing parameterization of the wave-induced melting of ice cliffs that considers the flow physics adjacent to the ice. Their parameterization of the melt rate relies on an empirical correlation between the friction coefficient and the local Reynolds number for oscillatory flow and on relating that friction coefficient to the heat transfer coefficient via an empirical relationship between the momentum diffusivity and the thermal diffusivity \citep[the so-called Reynolds analogy, \textit{e.g.},][]{Kundu2012}. This parameterization has been used by \cite{EL-Tahan:1987tk}, \cite{crawford2024evaluating}, and recently further developed by \cite{mamer2025buoyancy}, to estimate iceberg decay rates, and has also been mentioned in a number of studies since \cite[e.g.,][]{Savage:2001hz,Kubat:2007wi}. However, apart from other simplifications, it does not explicitly consider the unsteady, oscillatory boundary layer dynamics and how heat and momentum is transported within in it. As \citet{Weiss25} recently point out with field data, the unsteadiness of the boundary layer flow can cause the melt rates to be severely under-predicted by parameterizations that assume steady flow. On the other hand, field estimates of the melt rate in \cite{martin1978} and \cite{keys1984rates} are much smaller than the values calculated with the \citeauthor{White:1980up}'s parameterization, calling into question its applicability across different field conditions.

In terms of laboratory experimental data, there is scant previous work. \cite{White:1980up} report two wave-induced ice melting experiments in relatively warm freshwater, but without waterline melt rates. \citet{daley_veitch_2000} also report several experiments, but with insufficient detail to infer the precise conditions and melt rates. \cite{martin1978} reported an experiment with waves generated of period 0.4 s and amplitude 2.5 cm in a wave tank with a water depth of 40 cm at a uniform oceanic salinity (34 psu) and low ambient temperature ($4$\C). They measured a wave-induced notch that extended approximately 4 cm above the waterline, 11 cm below, and 8 cm into the ice block after 45 minutes. 

Based on this review of the literature, we can conclude that existing representations of wave-induced sidewall erosion in models are largely based on back-of-the-envelope scalings and scarce field and laboratory observations. So, in order to constrain the broad question of how waves erode ice cliffs, we consider an idealized setup from both theoretical and experimental points of view. Specifically, we present results from laboratory experiments and corresponding theoretical considerations that are made tractable by considering freshwater (ignoring the role of salinity) and non-breaking reflected waves. The findings presented here therefore do not represent a particular geophysical setting but rather are concerned with the fundamental interaction between waves and ice. We show that good agreement can be achieved between laboratory experiments and theoretical predictions, suggesting that the fundamental processes we identify do indeed play a key role in setting real-world erosion rates. 

In the remainder of the paper, we first present our theory of wave-induced melt of ice cliffs, including a new formulation for the predicted melt rate (\S\ref{sec:theory}), followed by laboratory experiments and their results (\S\ref{sec:labexpts}) and the implications of our work for geophysical and climate models (\S\ref{sec:implications}). We end with brief conclusions and directions for future work (\S\ref{sec:conclusions}).

\section{Theory}
\label{sec:theory}

\subsection{Standing waves at an ice cliff}
\label{sec:standingwaves}

\begin{figure}
  \centerline{\includegraphics[width = 0.8 \textwidth]{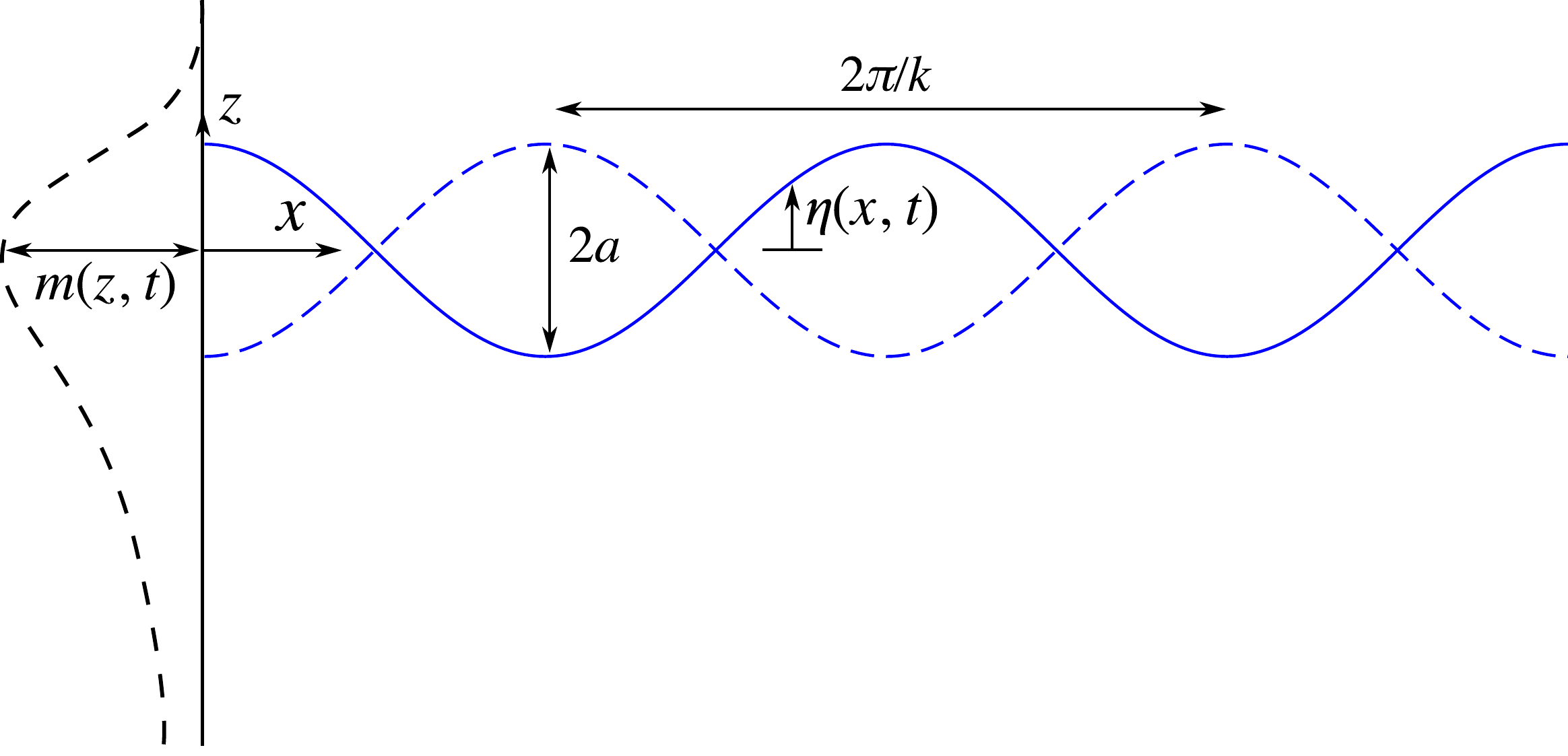}}
  \caption{Definition sketch for non-breaking surface waves reflecting at an ice wall. The wall is initially located at $x=0$, with the mean water level at $z=0$.}
\label{fig:definitionsketch}
\end{figure}

For planar, deep-water, small-amplitude surface gravity waves reflecting off a vertical ice wall (figure \ref{fig:definitionsketch}), the inviscid and irrotational solutions to the free surface displacement and velocity field can be constructed from a superposition of progressive waves traveling in opposite directions \citep[\textit{e.g.},][]{DeanDalrymple91, Mei05}. This solution is given by the real parts of 
\begin{subequations}
    \begin{align}
        \phi &= ig\frac{2a}{\omega}e^{kz}\cos kx\,e^{i\omega t}, \\
        \eta &= -\frac{1}{g}\frac{\partial\phi}{\partial t}\bigg|_{z=0}=2a\cos kx\,e^{i\omega t},\\
        u_I &= \frac{\partial\phi}{\partial x}=-ig\frac{2ka}{\omega}e^{kz}\sin kx\,e^{i\omega t}, \\
        w_I &= \frac{\partial\phi}{\partial z}=ig\frac{2ka}{\omega}e^{kz}\cos kx\,e^{i\omega t},
    \end{align}
\label{eq:deepwater_dimensional_exp}%
\end{subequations}
where $\phi$ is the velocity potential, $\eta$ is the free-surface displacement, and $u_I,w_I$ are the horizontal and vertical velocities, respectively. The subscript $I$ emphasizes that this is the inviscid and irrotational flow solution. The waves are characterized by angular frequency $\omega$ and wavenumber $k$, which follow the deep-water dispersion relation $\omega^2 =gk$ where $g$ is the gravitational acceleration. The incident waves have an amplitude $a$, which means the free surface oscillates between $-2a$ and $+2a$ at the anti-nodes in pure standing waves. 

\subsection{Wave-averaged heat transport in the ice-adjacent boundary layer}
\label{sec:boundarylayer}

To develop a theory of wave-induced ice melt, we use the passive scalar heat transport equation 
\begin{equation}
\frac{\partial\theta}{\partial t}+u\frac{\partial\theta}{\partial x}+w\frac{\partial\theta}{\partial z}=\alpha\bigg(\frac{\partial^2\theta}{\partial x^2}+\frac{\partial^2\theta}{\partial z^2}\bigg)
    \label{eq:dimensional_heat_ADE}
\end{equation}
where $\theta$ is the temperature, $(u,w)$ is the fluid velocity field, and $\alpha$ is the thermal diffusivity of the fluid. By treating temperature as a passive scalar, we ignore the feedback between temperature and fluid density that would result in buoyancy-driven natural convection flows.  We neglect these effects for the sake of simplicity in order to focus on the wave-induced heat transport and melt, and then evaluate their importance \textit{a posteriori} in laboratory data (\S\ref{sec:labexpts}).

In principle, \eqref{eq:dimensional_heat_ADE} can be solved using \eqref{eq:deepwater_dimensional_exp} with appropriate temperature boundary conditions. However, the wave-induced erosion of the ice will be strongly influenced by the dynamics and transport in the boundary layer near the ice--water interface. Furthermore, the flow described by the inviscid and irrotational solution in \eqref{eq:deepwater_dimensional_exp} is only a good description of the wave-driven flow outside the boundary layer, since the flow within boundary layer is affected by viscosity and includes rotational velocities. 

Oscillatory boundary layer flow is a classical problem in fluid mechanics, and the solution to the boundary layer flow due the reflection of surface waves from a vertical wall can be readily obtained from canonical analogous solutions \citep{Kundu2012, Batchelor00}. These solutions show that the leading order flow field in boundary layers driven by oscillatory flow remains oscillatory albeit modified by viscous effects. In terms of scalar transport, this oscillatory flow only serves to cause oscillatory advection with no net transport. However, oscillatory boundary layers can also produce Lagrangian mass transport at second order that can serve to cause a net advection of passive scalars \citep{Rayleigh1884, Batchelor00}. In particular, it has been shown that the boundary layer flow of surface waves interacting with a wall produces a wall-parallel Eulerian streaming current \citep{HuntJohns63, Mei05}, and that such a current is an important component of wave-averaged scalar transport dynamics where the effects of wave-induced oscillations have been averaged out \citep{Mei94, Winckler13, Michele21, Michele23}. 

Since the boundary layer flow and the corresponding passive scalar transport equation for surface waves reflecting at a vertical wall do not appear to have been previously published, we give a full derivation in Appendix~\ref{app:waveaveraged}. To summarize the appendix here, we use a multi-timescale expansion to isolate the slow, wave-averaged dynamics from the fast, wave-resolved dynamics. This procedure yields a wave-averaged advection-diffusion equation for the temperature, where the advection consists of both Eulerian currents and Stokes drift terms and where the horizontal diffusion dominates over the vertical because of the large gradients in the wall-normal direction across the boundary layer. For wave-induced melt of an ice cliff, we also show how the same multi-timescale expansion can be used to derive a wave-averaged Stefan condition that can be used to reduce the wave-averaged transport equation to a boundary value problem. Solving this boundary value problem gives the wave-averaged temperature field which can then be inserted back into the wave-averaged Stefan condition to compute the wave-averaged melt rate. 

Here, we only solve a simplified, approximate version of the full problem derived in the appendix. Specifically, we focus on how the wave-averaged temperature field is controlled by the balance between vertical advection by the Eulerian streaming current and horizontal diffusion. This simplified balance can be written as 
\begin{equation}
    W \frac{\partial\Theta}{\partial z} = \alpha \frac{\partial^2\Theta}{\partial x^2},
    \label{eq:WA-ADE-D1}
\end{equation}
where the capital variables $W$ and $\Theta$ signify wave-averaged versions of $w$ and $\theta$, respectively. For $W$, we use \citeauthor{{Batchelor00}}'s (\citeyear{Batchelor00}) expression for the Eulerian streaming current at the edge of the boundary layer that bypasses the need for a full multi-timescale analysis (see his Eq.~5.13.20). For our case, \citeauthor{{Batchelor00}}'s expression simplifies to
\begin{equation}
    W = -\frac{3}{8\omega} \frac{d ( \, w_I(x=0) w_I^*(x=0) \, )}{dz} = - 3 \frac{\omega}{k} (ka)^2 e^{2kz}, 
\end{equation}
where the $^*$ denotes a complex conjugate. This expression, which is consistent with our full solution in Appendix~\ref{app:waveaveraged}, is independent of the viscosity, which means that it would also remain valid in turbulent conditions under the assumption of a constant turbulent viscosity. 

Before presenting the solution, we first summarize the problem in dimensionless terms. We introduce the dimensionless variables
\begin{equation}
\begin{aligned}
    x \rightarrow \delta \frac{1}{k}\xi, \, z \rightarrow \frac{1}{k}z, W \rightarrow \frac{\omega}{k}W, \, \Theta \rightarrow (\theta_w-\theta_m)\Theta + \theta_m ,
\end{aligned}
    \label{eq:scaling_NS_ADE}
\end{equation}
where $\theta_m$ is the ice melting temperature and $\theta_w$ is the ambient water temperature. We have introduced a stretched coordinate $\xi$ for the wall-normal horizontal direction based on the boundary layer thickness, which is expected to be $O(\sqrt{\nu/\omega})$. Therefore, $\delta = \sqrt{k^2\nu/\omega}$ is the dimensionless boundary layer thickness. The equation to solve then becomes 
\begin{equation}
W \frac{\partial\Theta}{\partial z} = \mathrm{Pr}^{-1}\frac{\partial^2\Theta}{\partial \xi^2},
\label{eq:WA-ADE-R2}
\end{equation}
where $W =  - 3 \varepsilon^2 e^{2z}$ is the dimensionless advection velocity, with $\varepsilon = ka$ being the wave steepness and $\Pr = \nu/\alpha$ being the Prandtl number. The boundary conditions are 
\begin{subequations}
    \begin{align}
       \Theta(\xi=0) &= 0, \label{eq:tempBC-zero-WA} \\
       \Theta(\xi \rightarrow \infty) &= 1, \label{eq:tempBC-infty-WA} \\
        \frac{dM}{dT} &= - \mathrm{Ste} \, \mathrm{Pr}^{-1} \frac{\partial \Theta}{\partial \xi} \bigg\lvert_{\xi=0}, \label{eq:StefanBC-WA}
    \end{align}
    \label{eq:NS_ADE_BCs1}%
\end{subequations}
where $M$ is the wave-averaged position of the ice--water interface and $\mathrm{Ste} = c_p \lvert \theta_w - \theta_m \lvert /L$ is the Stefan number, with $c_p$ being the specific heat capacity of the water and $L$ being the latent heat of fusion of the ice. While \eqref{eq:tempBC-zero-WA} and \eqref{eq:tempBC-infty-WA} specify that the wave-averaged temperature must equal the melting temperature at the ice wall and recover to the ambient water temperature far outside the boundary layer, \eqref{eq:StefanBC-WA} is the wave-averaged Stefan condition (derived in Appendix~\ref{app:waveaveraged}) that sets the melt rate.

In sum, we must solve \eqref{eq:WA-ADE-R2} subject to boundary conditions \eqref{eq:tempBC-zero-WA} and \eqref{eq:tempBC-infty-WA} to obtain the wave-averaged temperature solution, and then use this solution to evaluate the wave-averaged melt rate using \eqref{eq:StefanBC-WA}. The relevant dimensionless parameters are the wave steepness $\varepsilon = ka$, the dimensionless boundary layer thickness $\delta = \sqrt{k^2\nu/\omega}$, the Prandtl number $\Pr = \nu/\alpha$, and the Stefan number $\mathrm{Ste}$, all of which tend to be small (see Appendix~\ref{app:waveaveraged} for more details).

\subsection{Wave-averaged temperature field and melt rate}

Physically, \eqref{eq:WA-ADE-R2} states that a downward advection of temperature whose strength scales with the square of the wave steepness and decays exponentially with depth must balance the horizontal diffusion. The solution to such a problem can be found in terms of similarity variables, analogous to cannonical boundary layer solutions (\textit{e.g.}, Blasius, Falkner-Skan).

First, we introducing a new vertical coordinate
\[
\zeta = e^z,
\]
so that the advection velocity becomes $-3\varepsilon^2 \zeta^{2}$ and $\partial / \partial z \rightarrow \zeta \partial /\partial \zeta$. Now, the equation to solve is
\begin{equation}
-3\varepsilon^2 \zeta^{3} \frac{\partial\Theta}{\partial \zeta} = \mathrm{Pr}^{-1}\frac{\partial^2\Theta}{\partial \xi^2}.
\label{eq:WA-ADE-Rb}
\end{equation}

We pose the similarity solution
\begin{equation}
\Theta = F(\chi), \text{ where } \chi = \frac{\xi}{\Delta(\zeta)}
\label{eq:similarity}
\end{equation}
is the similarity variable. Here, $\Delta(\zeta)$ corresponds to the thermal boundary layer thickness. Inserting \eqref{eq:similarity} into \eqref{eq:WA-ADE-Rb} gives
\begin{equation}
3\varepsilon^2 \zeta^{3} F^{\prime} \chi \frac{\Delta^{\prime}}{\Delta} = \mathrm{Pr}^{-1}\frac{F^{\prime \prime}}{\Delta^2},
\end{equation}
where the prime denotes differentiation. The introduction of the similarity variable has turned the partial differential equation into an ordinary differential equation, allowing us to rearrange the equation to have all terms that are functions of $\chi$ on the left side and all terms that are functions of $\zeta$ on the right side. Then, since $\chi$ and $\zeta$ are independent, each side must in turn be equal to a constant. Doing so gives
\begin{equation}
\frac{1}{3\varepsilon^2 \mathrm{Pr}} \frac{F^{\prime \prime}}{F^{\prime}} \frac{1}{\chi} = \Delta^{\prime} \Delta \zeta^3 = -C,
\end{equation}
where the negative sign of $C$ comes from inspection of the sign of $\Delta^{\prime}$. The thickness of the temperature boundary layer must increase with depth below the surface since the advective velocity decreases in magnitude with depth.

We first tackle the right side of the equation to find $\Delta$ and uncover the form of the similarity variable. We find that it can be written as
\[
\frac{d \left( \tfrac{1}{2}\Delta^2 \right)}{d\zeta} = -\frac{C}{\zeta^3},
\]
which can be integrated to give
\[
\Delta^2 = C\zeta^{-2},
\]
or with a choice of $C=1$,
\[
\Delta = \zeta^{-1}.
\]
This means that $\Delta  = e^{-z}$, which shows that the thermal boundary layer thickness increases exponentially with depth to match the exponential decrease in the magnitude of the wave-induced velocity. It also means that the final form of the similarity variable is
\begin{equation}
\chi = \xi \zeta
\end{equation}

We now turn to the left side of the equation to find the solution for $F$. With the choice of $C=1$ already made, we must solve the ordinary differential equation
\[
F^{\prime \prime} + 3 \varepsilon^2 \mathrm{Pr} \, \chi F^{\prime} = 0.
\]
We let $f = F^{\prime}$ to rewrite this equation as
\[
f^{\prime} + 3 \varepsilon^2 \mathrm{Pr} \, \chi f = 0.
\]
This is a first-order ordinary differential equation that can be solved with an integrating factor, producing the following solution
\[
\exp{\left( \frac{3}{2} \varepsilon^2 \mathrm{Pr} \, \chi^2 \right)}f = C_1,
\]
where $C_1$ is a constant of integration. Inserting $f = F^{\prime}$ then gives
\[
F^{\prime} = C_1 \exp{\left(- \frac{3}{2} \varepsilon^2 \mathrm{Pr} \, \chi^2 \right)},
\]
whose integral can be written in terms of error functions as
\[
F = C_1 \left[ \frac{\sqrt{\pi}}{2} \frac{ \mathrm{erf} \left( \sqrt{ (3/2) \mathrm{Pr} } \, \varepsilon  \chi \right) }{\sqrt{(3/2) \mathrm{Pr} } \, \varepsilon \, } \right] + C_2,
\]
where $C_2$ is another constant of integration.

\begin{figure}
  \centerline{\includegraphics[width = 0.65 \textwidth]{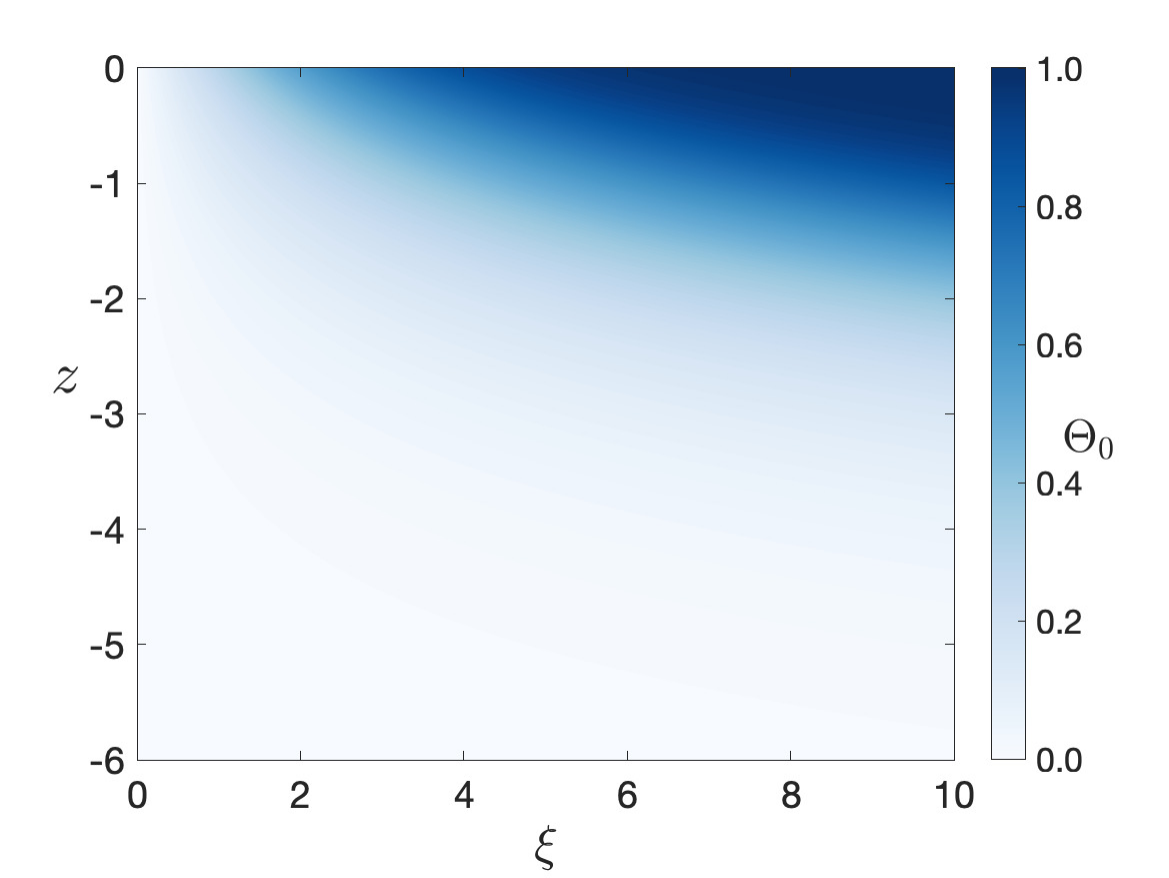}}
  \caption{Wave-averaged temperature $\Theta_0$ plotted against horizontal distance away from the ice--water interface $\xi$ and depth below the water surface $-z$ for wave steepness $\varepsilon =0.25$ and Prandtl number $\Pr = 5$ using \eqref{eq:tempsoln}.}
\label{fig:WAtemperature}
\end{figure}

The boundary conditions in terms of the similarity variables become $F(\chi = 0) = 0$ and $F(\chi \rightarrow \infty) = 1$. From the first boundary condition, we get $C_2 = 0$. From the second boundary condition, we find
\[
C_1 = \frac{2}{\sqrt{\pi}}\sqrt{\frac{3}{2} \mathrm{Pr} } \, \varepsilon.
\]
Thus, the final similarity solution is
\[
F = \mathrm{erf}\left( \sqrt{\frac{3}{2} \mathrm{Pr} } \, \varepsilon \chi \right),
\]
which when written in terms of the original dimensionless variables is
\begin{equation}
\Theta = \mathrm{erf} \left(  \sqrt{\frac{3}{2} \mathrm{Pr} } \,\, \varepsilon \, \xi \, e^{z} \right).
\label{eq:tempsoln}
\end{equation}
This solution for the wave-averaged temperature field is plotted in figure \ref{fig:WAtemperature}. It shows how the downward advection draws the warmer water closer to the ice, with this effect being strongest at the surface and decaying with depth.

Inserting \eqref{eq:tempsoln} into \eqref{eq:StefanBC-WA} gives the dimensionless wave-induced melt rate profile
\begin{equation}
\frac{dM}{dT} = \sqrt{\frac{6}{\pi}} \,\mathrm{Ste} \, \mathrm{Pr}^{-1/2} \, \varepsilon \, e^z,
\label{eq:meltrate}
\end{equation}
where we have dropped the negative sign with the understanding that this melt rate indicates that rate at which the the ice--water interface moves into the ice.

In dimensional form, the melt rate profile is given by
\begin{equation}
\frac{dm}{dt} = \sqrt{\alpha \omega } \, \sqrt{ \frac{6}{\pi} } \, \frac{c_p \lvert \theta_w - \theta_m \lvert }{L} \, ka \, e^{kz} = \left( \sqrt{ \frac{6 \alpha }{\pi}} \, \frac{c_p g}{L} \right) \, a \, \omega^{5/2} \, \lvert \theta_w - \theta_m \rvert \, e^{kz},
\label{eq:meltrate_dim}
\end{equation}
where the first expression emphasizes the melt rate scaling and the second expression isolates constants and material properties from the relevant environmental conditions. At the waterline ($z=0$), the melt rate is given by
\begin{equation}
\frac{dm}{dt} \bigg\rvert_{z=0}  = \left( \sqrt{ \frac{6 \alpha }{\pi}} \, \frac{c_p g}{L} \right) \, a \, \omega^{5/2} \, \lvert \theta_w - \theta_m \rvert.
\label{eq:meltrate_dim_waterline}
\end{equation}

\subsection{Interpretation of theoretical results and comparison against \citet{White:1980up}}
\label{sec:interpretationstheory}

In our theory, the near-ice temperature field is set by a balance of vertical heat transport by boundary layer streaming and horizontal heat transport by diffusion. Since the streaming velocity decays with the depth, the thermal boundary layer thickens with depth, and this sets the structure of the temperature field (figure \ref{fig:WAtemperature}). We have used the vertical streaming velocity at the edge of the boundary layer, which in dimensionless form is $W = -3 \varepsilon^2 \exp(2z)$ and in dimensional form is $W = -3 (a^2 \omega^3/g) \exp(2\omega^2 z/g)$. Since the streaming velocity at the boundary layer edge is independent of viscosity, the temperature solution and its implied melt rate are also valid in turbulent conditions under the assumption of a constant turbulent viscosity. 

We have found the wave-induced melt rate to scale as $\sqrt{\alpha \omega}$, which when written as $\omega \sqrt{\alpha/ \omega}$ can be interpreted as the speed needed to traverse the thermal boundary layer in one wave period. Further, the melt rate is linear with respect to both the wave steepness and the Stefan number, which in dimensional terms means that it is linear with respect to the incident wave amplitude and with respect to the temperature difference between the ambient water and the ice melting point. We also see that the melt rate decreases exponentially with depth with a decay length scale given by the inverse wavenumber. While this follows the same depth dependency as the wave-induced fluid velocities, the depth dependency in our melt rate actually comes from the fact that the streaming velocity decays exponentially with a decay scale of half the inverse wavenumber and the thermal boundary layer thickness scales as the square root of the streaming velocity. Apart from the exponential decay with depth, the only other non-linear dependency in our solution is that the melt rate follows a power law with respect to the wave frequency with an exponent of $5/2$. This means that, for a given wave amplitude or steepness, high frequency waves induce faster melting, but this melting will be confined to shallower depths since their wavenumber will be larger. Lastly, the melt rate also depends on a number of material properties, which themselves are functions of the water temperature and therefore can provide an additional non-linear dependency. However, the changes in the thermal diffusivity and specific heat capacity of water over realistic temperature ranges are typically small enough to neglect.

In contrast to our theory, \cite{White:1980up} proposed that the main mechanism of wave-induced melt of ice cliffs is the enhanced heat transfer caused by the oscillatory flow at an ice wall. They quantified the melt rate for this proposed mechanism using empirical correlations for the friction coefficient, which were linked to the heat transfer using another empirical correlation that employed a Reynolds analogy \citep[\textit{e.g.}][who define Reynolds analogy as the idea of equating thermal diffusivity with momentum diffusivity in turbulent flows]{Kundu2012}. Their resulting melt rate expressions are
\begin{subequations}
\begin{align}
    \frac{\id m}{\id t} &= \left\{ 5.04 \times 10^{-5} \left( \frac{a^2 \omega}{\nu} \right)^{-0.12} \right\} a\omega  \lvert \theta_w - \theta_m \rvert e^{kz} , \label{eq:W80_smooth} \\
    \frac{\id m}{\id t} &= \left\{ 4.05 \times 10^{-5} \left( \frac{r}{a} \right)^{0.2} \right\}  a\omega  \lvert \theta_w - \theta_m \rvert e^{kz}, \label{eq:W80_rough}
\end{align} \label{eq:W80} %
\end{subequations}  
where \eqref{eq:W80_smooth} is for a smooth ice wall and \eqref{eq:W80_rough} is for a rough ice all with $r$ being the roughness scale, and the terms in the square brackets coming from the various empirical correlations. 

Comparing \eqref{eq:W80} with our result \eqref{eq:meltrate_dim}, we see the same linear dependency with respect to the temperature $\lvert \theta_w - \theta_m \rvert$, but we see that the melt rates in \eqref{eq:W80} are fundamentally linked to the wave-induced oscillatory flow velocity scale $a \omega$, whereas our theory shows that the melt rate scales as the velocity needed to traverse the thermal boundary layer in one wave period $\sqrt{\alpha \omega}$. The other major difference is the scaling with respect to the wave frequency. While \eqref{eq:W80} show linear (or near-linear when accounting for the empirical correlations) dependency on the wave angular frequency $\omega$, we find that the melt rate scales as $\omega^{5/2}$.

\section{Laboratory experiments}
\label{sec:labexpts}

\subsection{Experimental methods}

\begin{figure}
  \centerline{\includegraphics[width = 0.95 \textwidth]{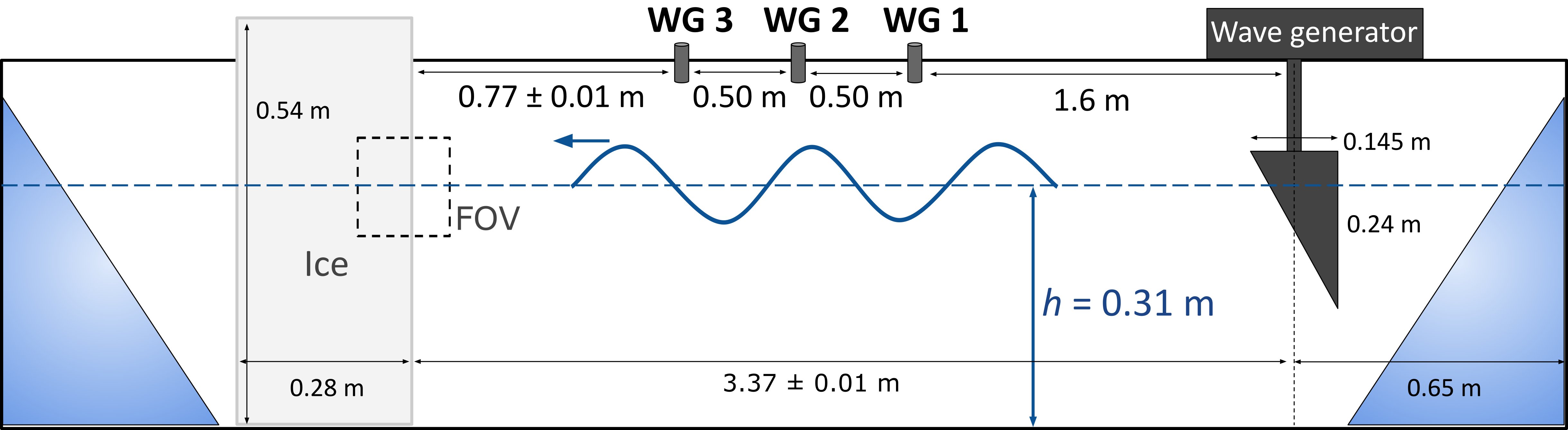}}
  \caption{Wave flume setup.}
\label{fig:flumesetup}
\end{figure}

We conducted laboratory experiments of wave erosion of ice cliffs using the setup shown in figure \ref{fig:flumesetup}, and which are summarized in table \ref{tab:expts}. We outline the experimental methods here, with more details available in \citet{Wolterman25}.

\begin{table}
    \begin{center}
    \begin{tabular}{ccc}
         ~ &\: $a$ (cm) &\: $\theta_w$ (\C) \\[3pt]
         W0T1 &\: 0 (no waves) &\: 21.2 $\pm$ 0.2 \\
         W1T1 &\: 0.35 $\pm$ 0.04  &\: 21.5 $\pm$ 0.2 \\
         W2T1 &\: 0.92 $\pm$ 0.04 &\: 21.7 $\pm$ 0.2  \\
         W3T1 &\: 1.55 $\pm$ 0.07 &\: 21.8 $\pm$ 0.2  \\
         W4T1 &\: 2.03 $\pm$ 0.07 &\: 22.0 $\pm$ 0.2  \\
         W3T2 &\: 1.55 $\pm$ 0.07 &\: 18.6 $\pm$ 0.3  \\
         W3T3 &\: 1.55 $\pm$ 0.07 &\: 13.3 $\pm$ 0.3 \\
    \end{tabular}
    \caption{Experiments performed at different incident wave amplitudes $a$ and ambient water temperatures $\theta_w$ with fixed water depth $h = 0.31$ m, wave angular frequency $\omega = 9.42$ rad s$^{-1}$, 
    salinity $S \approx 0$, air temperature $\theta_{\text{air}} \approx 22$\C, and ice surface temperature $\theta_{\text{ice}} \approx -5$\C.}
    \label{tab:expts}
    \end{center}
\end{table}

The wave flume of 5.0 m length, 0.21 m width and 0.51 m depth with glass walls was filled with fresh tap water to a depth of $h=0.31$ m. We generated waves using a triangular-wedge plunging-type wavemaker powered by a variable frequency motor (SureServo AC Servo Drive SVA-2100 controlled with a Tenma 72-8335A portable DC power supply). Wave-absorbing beaches made of coarse-pored aquarium sponge, cut into triangular prisms, were installed at both ends of the tank to absorb wave energy and limit wave reflection. Particle image velocimetry measurements of the fluid velocity field in this facility have previously shown this setup to produce high-quality surface waves \citep{BangPujara25}. 

In order to generate waves in the deep-water limit relevant for wave erosion of ice cliffs, we chose to operate the wavemaker near the highest frequency the system was capable of and varied the wave amplitude to produce different wave steepnesses. We collected data with three ultrasonic wave gauges (WGs; Senix ToughSonic 3 Distance sensors, accuracy 1 mm) spaced 0.5 m apart, which recorded water surface elevation time series at a sampling frequency of 50 Hz using a data acquisition system (NI USB-6002). Spurious data points were detected using thresholds and removed, and then the data were processed using a bandpass filter around the main input wave frequency to reduce small-scale noise. Using wave gauge data in experiments without ice to characterize the incident progressive waves, we determined the wave frequency by finding the peak of the free surface elevation spectra and found it to be consistently at an angular frequency $\omega = 9.42$ rad s$^{-1}$ (or $f = 1.5$ Hz). From this, we calculated the wavelength $2\pi/k = 0.69$ m and relative depth $kh = 2.8$ using the dispersion relation $\omega^2 = gk \tanh kh$. This relative depth is slightly smaller than the recommended limit for true deep-water conditions \citep[$kh \ge \pi$, according to][]{DeanDalrymple91} but large enough that the magnitude of the wave-induced flow at the flume bottom is a small fraction ($\approx 10\%$) of that at the free-surface. We determined the wave amplitude by computing the mean-square value of the free-surface elevation data, where the mean is taken as a moving average over one wave period. This value is $\tfrac{1}{2}a^2$ for sinusoidal functions, which allowed us to infer the wave amplitude $a$. 

To measure wave-induced erosion of ice cliffs, we placed a large ice block at the end of the flume opposite to the wavemaker for the various wave conditions listed in table \ref{tab:expts}. The ice blocks measured 0.28 m in length, 0.20 m in width, and 0.54 m in height, and were made with fresh tap water in coolers (Coleman Chiller, 48-quart capacity) kept in $-10$ to $-20$\C$\,$ freezers which pre-experiment were allowed to sit at room temperature until the ice surface warmed to $\approx -5$\C$\,$ (measured with a Traceable 4480 infrared thermometer, resolution $0.1$\C). When placed in the wave flume, the ice blocks spanned the full width and depth of the flume, held in place by their own weight and additional coarse-pored sponges as needed. The deterioration of the ice face was measured using quantitative imaging with a LaVision system (Imager MX 2M-160 1936$\times$1216 px camera mounted with a Tamron M112FM16 16 mm $f/2.0$ C-mount lens, controlled with DaVis 10.2). With the ice--water interface backlit using an LED panel and ambient light blocked with blackout material, we acquired 8-bit monochrome images with an average magnification factor of 6.4 px/mm at a sampling frequency of 10 Hz. In order to ensure reasonably high quality data, we positioned the ice blocks as close as possible to an integer number of wavelengths away from the wavemaker ($2\pi [\Delta x]_\text{ice-wavemaker}/k \approx$ 4.9) so that the wave conditions in the tank were close to purely standing waves. Further, we only measured melt rates during the period between when the first waves arrived at the ice face to when the width of the ice block decreased to approximately 70\% of the flume width so that the lateral (side-wall) contribution to the wave-induced melt rate was small. Finally, to quantify ice melting in the absence of waves, we also conducted an experiment where an ice block was allowed to melt in an otherwise quiescent environment at a water temperature of $\theta_w \approx 21$\C$\,$ (the no-waves case in table \ref{tab:expts}). We wanted to quantify the no-waves melting at this ambient water temperature since it represents a large thermal forcing, and we used the results of that test to understand the relative contributions of wave-induced melting and ambient melting at different wave amplitudes. 

We obtained measurements of the melt rate from the images using two different methods: (i) the ``submerged profile'' method, which focused on the vertical variation of the melt rate underwater and (ii) the ``waterline value'' method, which focused on the deepest (i.e. most eroded) part of the wave-cut notch near the mean free surface position. As we show in \S\ref{sec:exptresults}, the melt rates were reasonably consistent across the two methods.

(i) To obtain the submerged profile, we used the Canny method with MATLAB's edge detection function \textit{edge} followed by median filtering in space and then in time to remove spurious edge positions. We found the melt rate by computing the change in edge position divided by the change in time for all possible pairs of data points and then taking the median value to obtain a melt rate value at each pixel row that was robust to outliers. The melt rate profile data was then binned by depth and averaged across 1 cm vertical bins. We found it was particularly challenging to obtain clean data for edge positions (and hence melt rates) near the free surface due to light contamination and because of the projection of the free surface across the width of the flume onto the images. Hence, all data for the submerged profile are for $z \leq -1$ cm. Other challenges to pinpointing the edge position included ambiguities induced by the ice becoming more transparent as it thawed and the appearance of small-scale morphological features on the submerged melt face likely due to a transition to turbulence in the ice-adjacent boundary layer \citep[\textit{e.g.},][]{Josberger:1981gg, Bushuk19, PeglerWykes2020}. In particular, we observed that the submerged melt face was covered with small `scallops' in experiments with waves whereas `channels' formed in the deeper parts of the submerged melt face in experiments without waves, but neither were analyzed further.

(ii) To obtain the waterline value of the melt rate, we used the following steps. We first pre-processed the images with a two-dimensional median filter to remove small-scale noise followed by local contrast enhancement to improve edge detectability. Then, we used a simple threshold for intensity jump in each pixel row to identify the ice edge and removed any obvious spikes in this ice edge position data. We found the waterline melt rate by averaging the edge position data in the vicinity of the mean waterline ($-a \lesssim z \lesssim + a$) and fitting a straight line to this notch-averaged edge position time series. We found this method to be more robust to ambiguities of the ice edge position near the free surface that plagued the submerged profile method.

For both methods, we found that sub-sampling the images to one per wave period was the best compromise between temporal resolution (minimizing time between successive snapshots of the melt face) and accuracy (allowing enough time for the melt face to move a detectable amount). Generally we used images where the ice-adjacent free surface was at its maximum value so that almost all of the wave-induced erosion was underwater, but for the waterline value method we occasionally used images where the ice-adjacent free surface was at its minimum value so that the deepest part of the wave-cut notch was exposed to the air. 

\subsection{Experimental results and analysis}\label{sec:exptresults}

\begin{figure}
  \centerline{\includegraphics[width = 0.6 \textwidth]{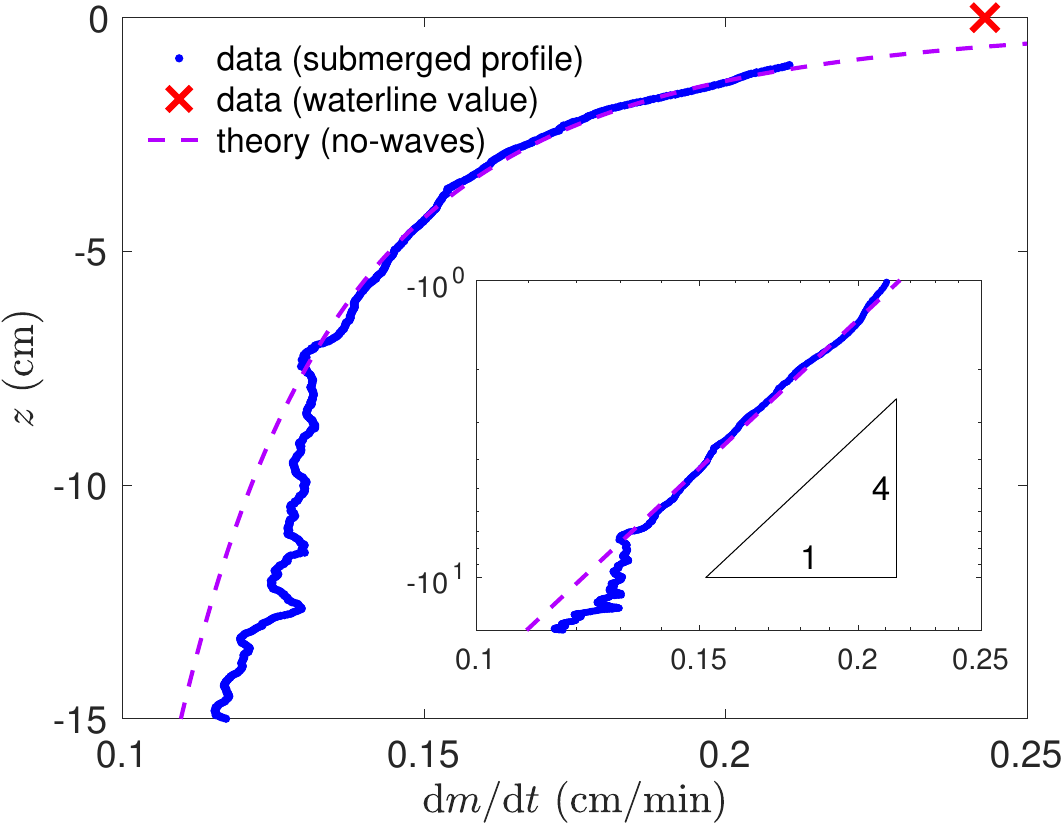}}
  \caption{Vertical profile of melt rate $\mathrm{d}m/\mathrm{d}t$ plotted against the vertical coordinate $z$ in an otherwise quiescent background environment at $\theta_w \approx 22$\C. The no-waves theory (dashed purple line) is \eqref{eq:nowaves_theory} with $A = 0.216$ cm$^{3/4}$/min. The inset is the same plot on logarithmic axes.}
\label{fig:TheoryData_NoWaves}
\end{figure}

We begin with an analysis of the no-waves melt rate data, which is shown in figure \ref{fig:TheoryData_NoWaves}. The $0$\C$\,$ melt water is expected to sink due to its negative buoyancy relative to the ambient $22$\C$\,$ water and create a natural convective laminar flow. The heat transfer between a vertical wall and the adjacent fluid in this configuration, as well as the resulting natural convection and melt, has been previously considered using boundary layer theory by \citet{Ostrach53, Acrivos60, PeglerWykes2020}. The result most relevant to this study is that the melt rate profile is given by
\begin{equation}
    \frac{dm}{dt} = A(-z)^{-1/4},
    \label{eq:nowaves_theory}
\end{equation}
where the parameter $A$ encapsulates all effects of material properties. Following \citet{PeglerWykes2020}, we empirically determine the constant $A$ from measurements by fitting the melt rate profile data over range $-7\ \mbox{cm}\leq z \leq - 1 \ \mbox{cm}$ to the power law in \eqref{eq:nowaves_theory} using $A$ as a fitting parameter. 

We observe that the submerged melt rate profile shows an excellent fit to the $(-z)^{-1/4}$ power law over the fitted range of depths, with small deviation and variations away from \eqref{eq:nowaves_theory} in the melt rate profile below this range. Very near the free surface, this theory predicts that the melt rate continues to increase and becomes infinite at $z=0$. In our data, we observe that this theoretical singularity is damped out, likely by effects not accounted for in the theory such as surface tension and a lack of fluid above this point to feed the convective flow. Our waterline value of the melt rate, which averages the deepest part of the waterline notch, is close to the theory value at $z \approx -0.6$ cm, suggesting that these near-surface effects are confined to top 1 cm or so.

\begin{figure}
  \begin{center}
  \subcaptionbox{\label{fig:TheoryData_H0T22}}{%
    \includegraphics[width=0.49\textwidth]{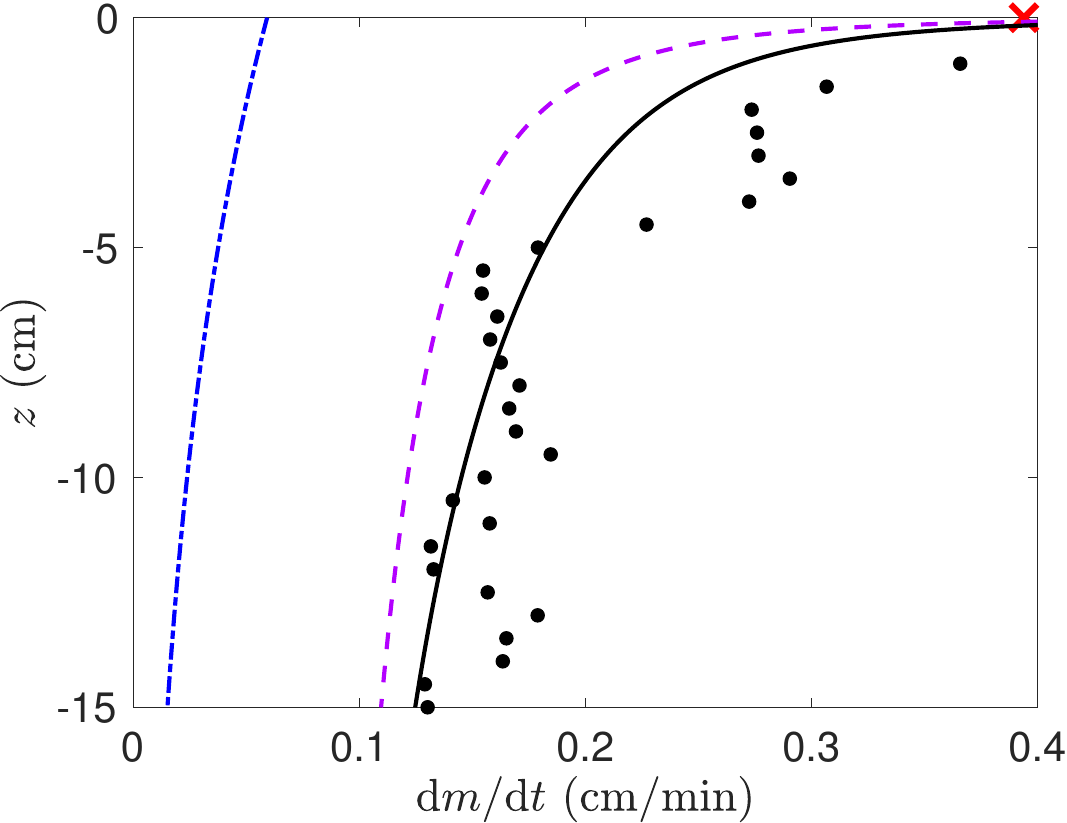}}
  \subcaptionbox{\label{fig:TheoryData_H1T22}}{%
    \includegraphics[width=0.49\textwidth]{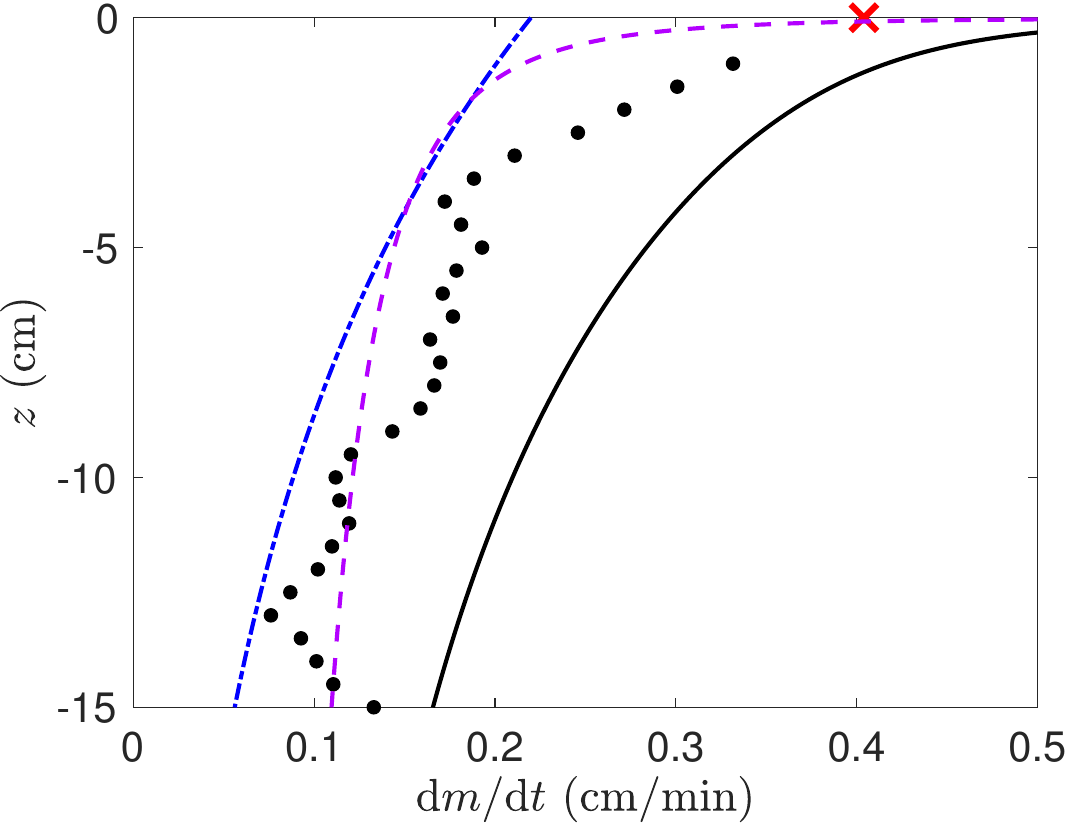}} \\
      \subcaptionbox{\label{fig:TheoryData_H2T22}}{%
    \includegraphics[width=0.49\textwidth]{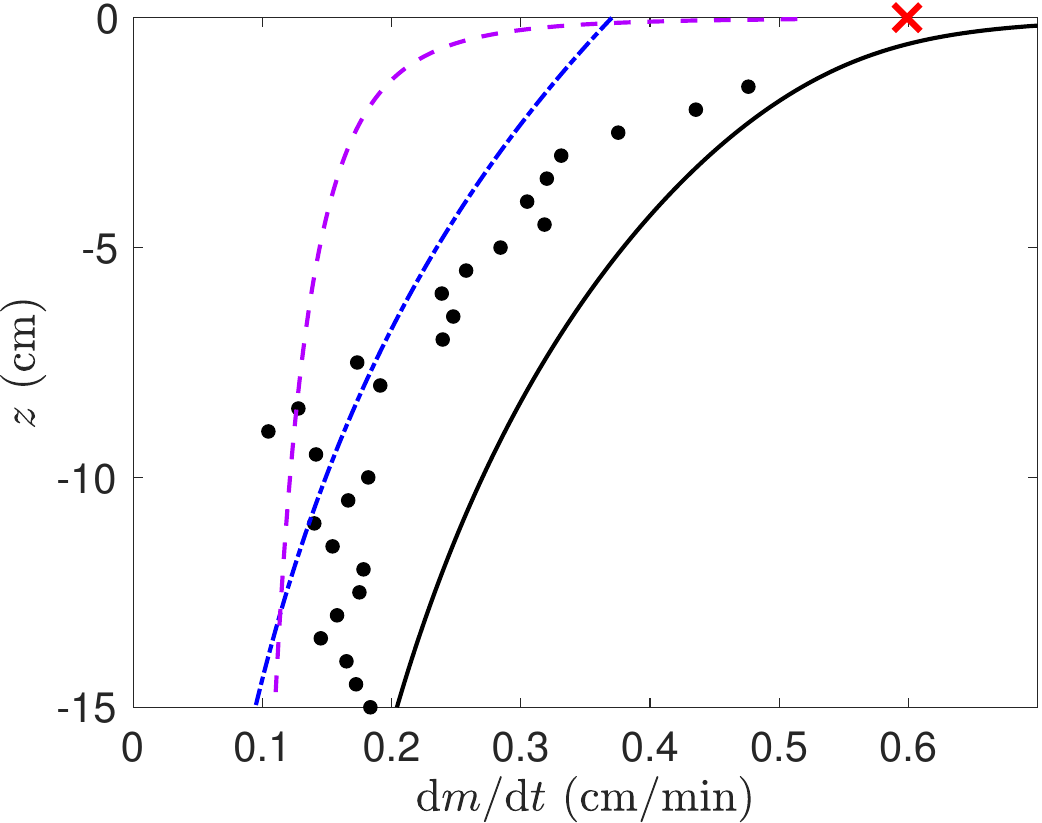}}
  \subcaptionbox{\label{fig:TheoryData_H3T22}}{%
    \includegraphics[width=0.49\textwidth]{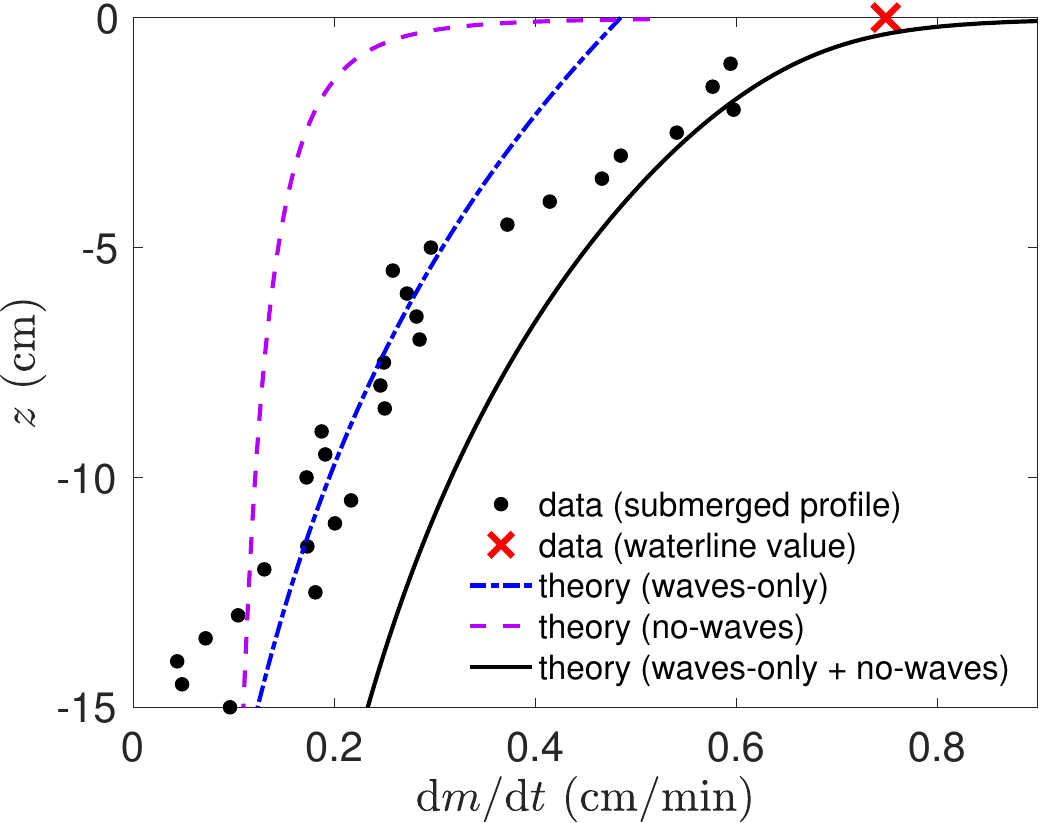}}
  \end{center} 
  \caption{Vertical profile of melt rate $\mathrm{d}m/\mathrm{d}t$ plotted against the vertical coordinate $z$ at $\theta_w = 22$\C$\,$ for different wave amplitudes: (a) $a=0.35$ cm; (b) $a=0.92$ cm; (c) $a=1.55$ cm; (d) $a=2.03$ cm. The no-waves theory (dashed purple line) is \eqref{eq:nowaves_theory} with $A = 0.216$ cm$^{5/4}$/min, the waves-only theory (dashed-dotted blue line) is \eqref{eq:meltrate_dim}, and the combined theory (thick black line) is the sum of the no-waves and waves-only lines.}
  \label{fig:TheoryData_HT22}
\end{figure}

We now turn to the main results in figure \ref{fig:TheoryData_HT22}, which show the melt rates for four different wave amplitudes (0.35 cm $\leq a \leq$ 2.03 cm) at the same wave frequency ($\omega = 9.42$ rad/s) and water temperature ($\theta_w \approx 22$\C). The submerged melt rate profile and the waterline melt rate value are compared against the  melt rate predictions for quiescent conditions (no-waves theory; \eqref{eq:nowaves_theory}), for wavy conditions (waves-only theory; \eqref{eq:meltrate_dim}), and their sum. We note that summing the two melt rate theories assumes that they act independently of each other, which is likely not the case, but we take this approach here for simplicity in the absence of a more advanced theory that considers both effects simultaneously.

Overall, we observe that the theory is able explain the main variation in the data with respect to the vertical coordinate and with respect to the wave amplitude, especially given the many simplifications and the absence of fitting parameters. (While the value of $A$ in the no-waves theory \eqref{eq:nowaves_theory} was fit to data, it is technically parameterizing difficult-to-measure material properties, as noted in \citet{PeglerWykes2020}. The waves-only theory \eqref{eq:meltrate_dim} includes constants related to material properties explicitly, so there are no fitting parameters.) 

Considering the comparison of the data against the theory as a function of wave amplitude, we note several interesting points. As the wave amplitude increases, the melt rate data transitions from matching the no-waves theory curves to matching the waves-only theory curves. This is especially true at smaller depths ($z \gtrsim -10$ cm) where the influence of the vertical streaming velocity is expected to be important. To better understand this, we recall that the streaming velocity scales with the square of the wave amplitude ($W \sim a^2 \omega^3 / g$) and decays with depth at twice the rate of the wave-induced fluid velocities ($W \sim e^{2kz}$). For our wave conditions, this means that the streaming velocity is predicted to be larger by a factor of more than 30 for the largest waves compared to the smallest waves, and to decay by approximately $85$\% at a depth of $z = -10$ cm compared to its value at the mean free surface. Physically, this suggests that at small depths and for large waves, the wave-induced streaming current sets the near-ice temperature field as shown in figure \ref{fig:WAtemperature}. Otherwise, the near-ice temperature field and melt rate is dominated by the natural convection flow \citep{Ostrach53, Acrivos60, PeglerWykes2020}. 

However, the picture changes very near the surface ($z \gtrsim - 3$ cm in our data). The data in this region show the influence of the no-waves melt for all wave amplitudes. In particular, the melt rate data shows a sharp transition from an exponential dependency on depth (waves-only theory) to a power law dependency (no-waves theory). In the same vein, the waterline melt rates are much greater than the predictions of the waves-only theory. Thus, the near-surface melt rate profile is a mixture of exponential and power law proportionalities. This suggests that despite the oscillating nature of the ice-adjacent free surface in wavy conditions, the waterline melt rate is still strongly influenced by the natural convection dynamics that lead to fast melt rates near the free surface. Even for the largest waves in our experiments, the data show that waves-only theory is insufficient to predict the waterline melt rates at this high level of thermal forcing.  

\begin{figure}
  \centerline{\includegraphics[width = 0.6 \textwidth]{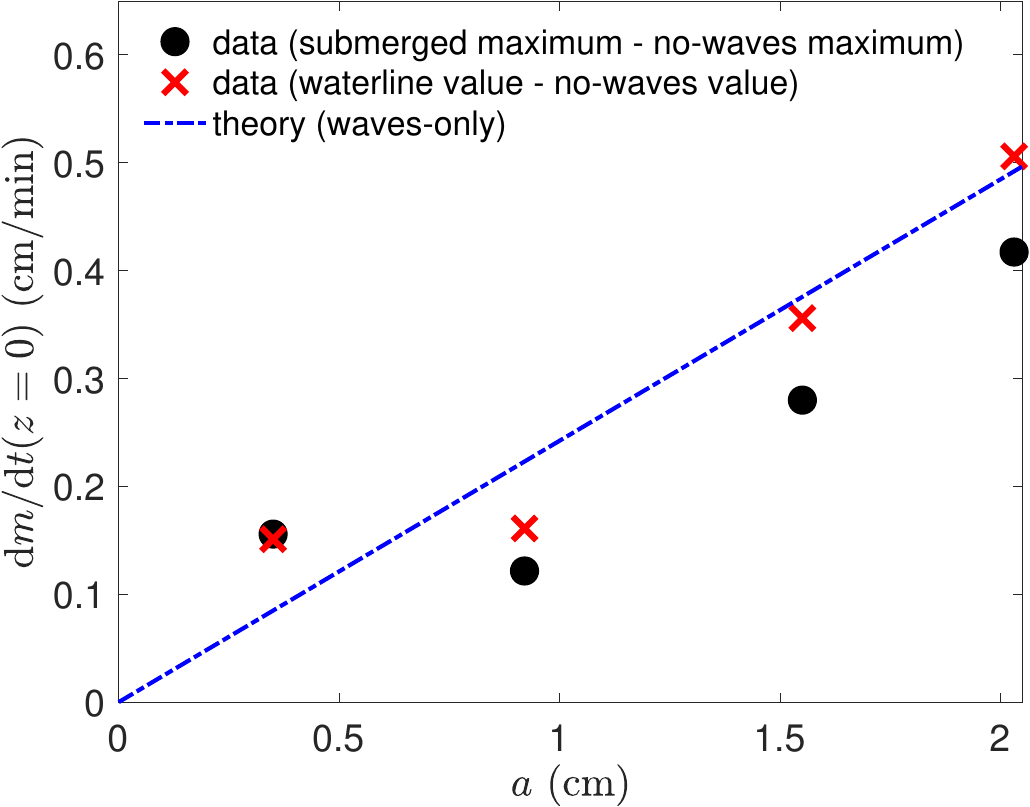}}
  \caption{Waterline melt rate $\mathrm{d}m/\mathrm{d}t$ (with the no-waves contribution subtracted) plotted against wave amplitude $a$ at $\theta_w = 22$\C. The waves-only theory (dashed-dotted blue line) is \eqref{eq:meltrate_dim_waterline}.}
\label{fig:TheoryData_T22_amplitudes}
\end{figure}

To further examine how well our waves-only theory performs at predicting the waterline melt rate, we show in figure \ref{fig:TheoryData_T22_amplitudes} the melt rate data as a function of wave amplitude with the no-waves melt rate subtracted off. We focus on the near-surface region by showing only the maximum melt rate observed in the submerged profile (typically at $z = -1$ cm) and the waterline melt rate value. This data is compared against the waves-only theory. We see that, despite the many simplifications, the data for the waterline melt rate agree with the theory remarkably well, especially as the wave amplitude increases.

\begin{figure}
  \begin{center}
  \subcaptionbox{\label{fig:TheoryData_H2T19}}{%
    \includegraphics[width=0.49\textwidth]{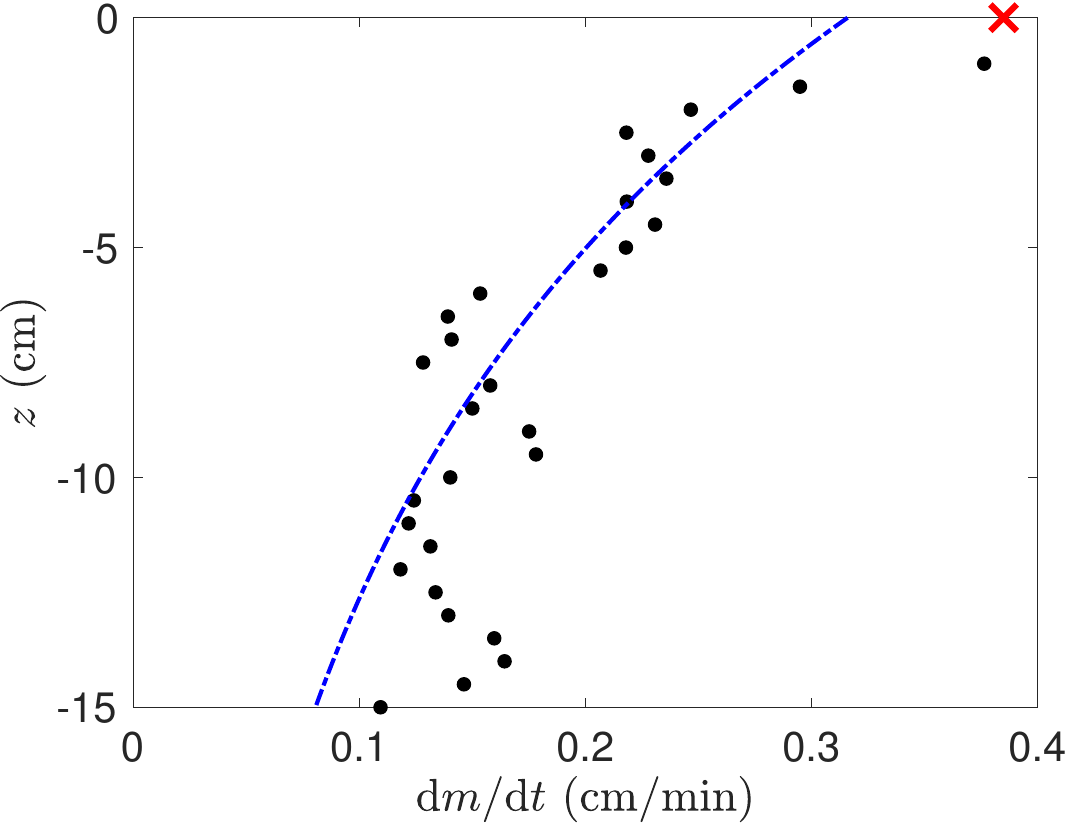}} 
      \subcaptionbox{\label{fig:TheoryData_H2T13}}{%
    \includegraphics[width=0.49\textwidth]{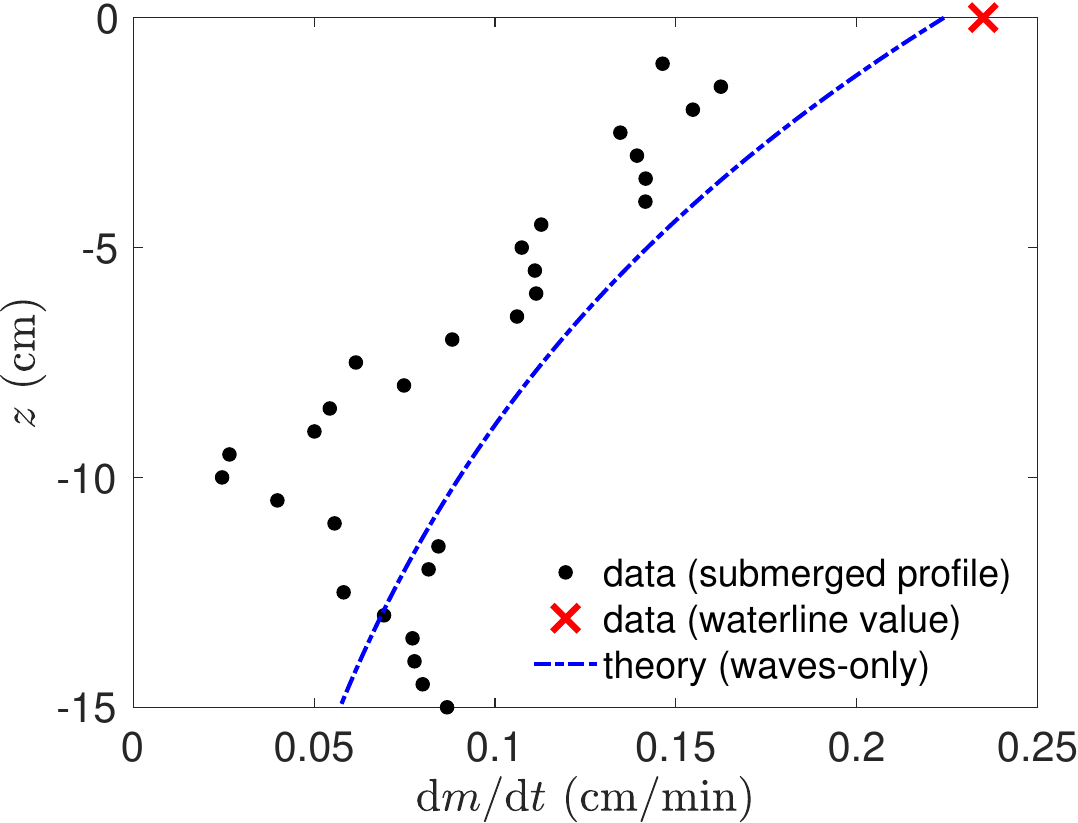}}
  \end{center} 
  \caption{Vertical profile of the melt rate $\mathrm{d}m/\mathrm{d}t$ plotted against the vertical coordinate $z$ at $a=0.1.55$ cm for different water temperatures: (a) $\theta_w = 18.6$\C; (b) $\theta_w = 13.3$\C.}
  \label{fig:TheoryData_H2T}
\end{figure}

Finally, we examine the melt rate profiles and waterline melt rate values for the colder water temperatures in figure \ref{fig:TheoryData_H2T}. At these lower ambient water temperatures that are more relevant to field conditions, the relative contribution of the no-waves melt with respect to the wave-induced melt should decrease. While we do not have no-waves melt data for these temperatures, the data confirm that the melt rates indeed match the waves-only theory much better when the ambient water is colder. For $\theta_w \approx 19$\C, the melt rate profile follows the waves-only theory quite well at depths $z \lesssim -2$ cm, and only very near the surface does the data deviate away from the waves-only theory. For the coldest water temperature $\theta_w \approx 13$\C, almost the entire melt rate profile, including the waterline melt rate value, follows the waves-only theory reasonably well, though we note that melt rates are overall smaller and the data are accordingly noisier compared to the warmer water temperature experiments.  

\section{Implications for geophysical and climate modeling}
\label{sec:implications}

Here, we discuss how wave erosion of ice cliffs has been represented in models to date, and how our results may help improve these representations.

Iceberg decay is driven by multiple processes \citep[\textit{e.g.},][]{Savage:2001hz, Cenedese2023}, but the wave-driven ice loss has long been regarded as the primary ablation mechanism for icebergs in the open ocean. As such, it is typically the largest contributor in iceberg models, representing up to approximately 80\% of total simulated iceberg mass loss \citep{EL-Tahan:1987tk, Bigg:1997bp, Martin:2010kb, Wagner:2017tc}. However, the majority of iceberg modules in current climate models use a parameterization for sidewall erosion based on the work of \cite{Bigg:1997bp}, which was initially conceived to roughly reproduce the lifespans of small icebergs in the Arctic. It has since been found that the melt rates computed in this way substantially underestimate the decay of many icebergs, particularly those of large tabular icebergs in the Southern Ocean \citep{Rackow:2017ej,bouhier2018melting,england2020modeling}. 

For glaciers and ice-shelf fronts, the traditional view is that environmental conditions are often such that direct wave-driven erosion is largely suppressed and ice loss is typically dominated by submarine melting and calving driven by other processes \cite[see][for a review]{benn2007calving}. And although calving due to enhanced melt rates at the waterline (potentially due to waves) is discussed in detail in \citeauthor{benn2007calving}'s (\citeyear{benn2007calving}) review, it has previously received little attention in the literature on marine-terminating glacier and ice-shelf deterioration---at least partly because it remains difficult and dangerous to collect field data at these ice cliffs. However, several recent studies have found that waterline erosion may indeed be an important driver of smaller-scale but frequent calving events at major ice discharge locations in both Greenland and Antarctica \citep[\textit{e.g.},][]{wagner2016role, slater2018localized, wagner2019large, becker2021buoyancy, sartore2025}. Further, for freshwater-terminating glaciers, waterline notch development has been shown to be an important driver of ice erosion  \cite[\textit{e.g.},][]{kirkbride1997calving, haresign2005melt, rohl2006thermo}.

As the preceding discussion illustrates, large scale models of icebergs, glacier fronts, and ice-shelf fronts are increasingly needing to include the effects of wave-induced erosion in waterline melt rates. In this regard, the most commonly used parameterization comes from \cite{Bigg:1997bp}, who introduced the expression for sidewall erosion $ \id m/ \id t = S_s/2 $, where $\id m/ \id t$ is the full-depth melt rate (in m/day) and $S_s$ is the Beaufort sea state, itself typically parameterized in terms of surface wind speed. \cite{Gladstone:2001cq} augmented this expression by adding a sea surface temperature dependency and an \textit{ad hoc} term accounting for the wave-damping effect of sea ice. Their expression reads $
\id m/\id t =S_s[1+\cos(C^3 \pi)] \lvert \theta_w - \theta_m \rvert/12$, where $C$ is the local sea ice concentration and the sea state is modeled following \citet{Martin:2010kb} via $S_s = a_1 |\boldsymbol{v}_a|^{1/2} + a_2 |\boldsymbol{v}_a|$, where $a_1$ and $a_2$ are tuning parameters and $\boldsymbol{v}_a$ is the surface wind speed. Thus, the wave erosion of ice cliffs in these models is fundamentally tied to the local wind, which ignores the impact of remotely generated ocean swells. Furthermore, even if the parameterization for the wave-induced waterline erosion gives accurate predictions, it is often applied as a full-depth melt rate.

Despite its coarseness, this type of parameterization has been widely adopted. For example, \cite{Marsh:2015dn} incorporated it into an iceberg module for the Nucleus for European Modeling of the Ocean (NEMO) framework, which was revised by \cite{Merino:2016jm} to include vertically variable melt rates, and different variations of it have been used in many other climate-related studies \cite[\textit{e.g.},][]{Jongma:2009cl,Jongma:2013hz,wilton2015modelling,BugelmayerBlaschek:2016gn,Stern:2016kh,Wagner:2017up,wagner2018wave,mackie2020climate,england2020modeling,huth2022parameterizing}. However, as mentioned above, it has also been shown that this parameterization leads to decay rates that are far too slow for large icebergs in the Southern Ocean, with the main culprit likely being insufficient wave-induced melting and calving. Additionally, when \cite{sartore2025} considered wave-induced erosion at the front of the Ross Ice Shelf with this parameterization, they found that it overestimated the melt rates by an order of magnitude compared to observations, likely due to the overemphasis on local wind speeds. The Ross Ice Shelf (and many others) experiences strong katabatic winds, but without sufficient fetch to create large waves. These points underscore that more physically grounded parameterizations of the wave-induced waterline melt would allow for more reliable representations of ice loss in geophysical and climate models, though part of the challenge is also clearly rooted in representing wave erosion without directly accounting for the wave conditions.

While the prevailing parameterization of \citet{White:1980up} given here in \eqref{eq:W80} has stood for several decades, we suggest that our new model given in \eqref{eq:meltrate_dim} may be better placed at capturing the dominant mechanism behind wave-cut notches into ice cliffs. Our laboratory results provide good support for this suggestion. While both formulations neglect salinity and temperature-related buoyancy effects, our calculations of the boundary layer flow and heat transport in Appendix~\ref{app:waveaveraged} suggest that the heat advection due to the wave-averaged Eulerian streaming current is of leading order importance and that the heat transport due to the wave-induced oscillatory flow is of secondary importance. The \citeauthor{White:1980up} parameterization neglects the role of the wave-averaged Eulerian streaming current, but not only is such a current is expected to occur, its magnitude is expected to be scale independent since it is independent of the viscosity. In other words, a similar current of a similar magnitude is also expected in turbulent conditions in a Reynolds-averaged sense, which means that \eqref{eq:meltrate_dim} can be applied in large-scale models without the need for any empirical coefficients. 

\section{Conclusions}
\label{sec:conclusions}

We have considered anew the problem of wave-induced erosion of ice cliffs. By examining the boundary layer flow and heat transport for the reflection of surface waves off a vertical wall, we identified a novel mechanism that could be what controls the wave-induced ice melt. This mechanism stems from the Eulerian streaming current that is generated by oscillatory boundary layers in the interaction of waves with walls. We solve for the wave-averaged temperature field by balancing the heat advection by this wall-parallel vertical current with the horizontal heat diffusion. With the use of the Stefan condition, this allows us to generate a new formulation for the wave-induced melt rate profile, given in dimensional form in \eqref{eq:meltrate_dim}. In contrast to previous efforts, our new formulation is free of any empirical constants or correlations. It predicts that the waterline melt rate scales as $a \omega^{5/2} \lvert \theta_w - \theta_m \rvert$, where the waves incident on the ice cliff hae an amplitude $a$ and angular frequency $\omega$ and $ \lvert \theta_w - \theta_m \rvert$ gives the thermal forcing in terms of the temperature difference between the ambient water temperature and the ice melting temperature.

We test the ability of our formulation to predict the melt rate of ice blocks subject to wave action in laboratory experiments conducted in a wave flume. We vary the wave amplitude and ambient water temperature at a fixed wave frequency and water depth. The comparison between the theory and data show good agreement, but also reveal the role of the ambient ice melt unrelated to the waves. As the wave amplitude increases and the thermal forcing decreases, the role of the ambient melt is diminished and  we can adequately correct for it by simply subtracting it off the measured melt rate to arrive at an estimated measurement of the melt rate only due to waves. In this case, the data provide remarkably unambiguous support for our formulation at the largest two wave heights tested.

To highlight the approximations and limitations, we note that future experiments should explicitly test the non-linear dependency on the wave frequency that was not possible in this study due to limitations of the wave generator. We also note that our theory uses a constant density for the fluid. This neglects how variations in salinity and temperature would lead to changes in density, and hence drive convection. The deviation of our theory due to salinity- and temperature-related buoyancy effects would likely depend on the relative magnitudes of vertical melt water velocity compared with the Eulerian streaming velocity, which in turn, depend on the environmental conditions, including the ambient water temperature and salinity, as well as the melt rate itself. Such effects may be able to be included in a more complex theoretical framework that can include heat transport due to a mixture of wave-induced flow and natural convection \citep[\textit{e.g.},][]{JosbergerMartin81, Ostrach53, KerrMcConnochie15,  Wykes2018, PeglerWykes2020, PeglerWykes21, mamer2025buoyancy}. 


\backsection[Supplementary data]{\label{SupMat}Supplementary material and movies are available at [to be filled upon acceptance]}

\backsection[Acknowledgements]{The authors wish to thank those that helped with the experimental setup (J. Y. Bang, J. Koseff, P. Sobol, A. Stephens, L. Sunberg, J. Zeuske), carrying out the experiments (T. Bailey, A. M. Mixtli, J. Prescott), and data analysis and interpretation (M. Mamer, A. Robel, G. Verhille). The authors also thank J.-L. Thiffeault for insightful discussions related to the theory. Finally, some figures used color maps from a public repository \citep{Bewer25}.}

\backsection[Funding]{This research was supported by the US National Science Foundation (OPP-2148544). N.P. acknowledges an Early-Career Research Fellowship from the Gulf Research Program of the National Academies of Science, Engineering, and Medicine. A.W. acknowledges support from the Roy F. Weston fellowship at UW-Madison.}

\backsection[Declaration of interests]{The authors report no conflict of interest.}

\backsection[Data availability statement]{The data that support the findings of this study are openly available in [repository name] at http://doi.org/[doi], reference number [reference number]. [To be filled upon acceptance] 
}

\backsection[Author ORCIDs]{A. Wolterman, https://orcid.org/0000-0003-4250-578X; T. J. W. Wagner, https://orcid.org/0000-0003-4572-1285; L. K. Zoet, https://orcid.org/0000-0002-9635-4051; N. Pujara, https://orcid.org/0000-0002-0274-4527}

\backsection[Author contributions]{N.P., A.W., L.K.Z., and T.J.W.W. designed the study, A.W. conducted the experiments, A.W. and N.P. performed data analysis, all authors contributed to the interpretation of results and the writing of the manuscript.}

\appendix

\section{Wave-averaged flow, heat transport, and melt rate}\label{app:waveaveraged}

In this appendix, we derive expressions for the wave-averaged flow and passive scalar transport in the boundary layer adjacent to a vertical wall that reflects surface gravity waves. We perform this calculation in by subjecting the Navier--Stokes and passive scalar transport equations, and their respective boundary conditions, to a multi-timescale expansion. This is done within the context of heat transport near and melting of an ice cliff subject to wave action, but the wave-averaged scalar transport equation result could also be used for other applications. Then, using the Stefan condition, we show how to compute the wave-averaged melt rate.

We start with the two-dimensional, time-dependent Navier--Stokes and heat transport equations, which in dimensional form are
\begin{subequations}
    \begin{align}
        \text{continuity}&:\qquad\frac{\partial u}{\partial x}+\frac{\partial w}{\partial z}=0, \\
        x\text{-momentum}&:\qquad\frac{\partial u}{\partial t}+u\frac{\partial u}{\partial x}+w\frac{\partial u}{\partial z}=-\frac{1}{\rho}\frac{\partial p}{\partial x}+\nu\bigg(\frac{\partial^2 u}{\partial x^2}+\frac{\partial^2 u}{\partial z^2}\bigg),\\
        z\text{-momentum}&:\qquad\frac{\partial w}{\partial t}+u\frac{\partial w}{\partial x}+w\frac{\partial w}{\partial z}=-\frac{1}{\rho}\frac{\partial p}{\partial z}+\nu\bigg(\frac{\partial^2 w}{\partial x^2}+\frac{\partial^2 w}{\partial z^2}\bigg)-g,\\
        \text{heat transport}&:\qquad\frac{\partial\theta}{\partial t}+u\frac{\partial\theta}{\partial x}+w\frac{\partial\theta}{\partial z}=\alpha\bigg(\frac{\partial^2\theta}{\partial x^2}+\frac{\partial^2\theta}{\partial z^2}\bigg),
    \end{align}
    \label{eq:dimensional_NS_ADE}%
\end{subequations}
where $(u,w)$ is the flow field, $\theta$ is the temperature, $p$ is the pressure, $g$ is the gravitational acceleration, $\rho$ is the density, $\nu$ is the kinematic viscosity, and $\alpha$ is the thermal diffusivity. 

We proceed by making \eqref{eq:dimensional_NS_ADE} dimensionless using
\begin{equation}
\begin{aligned}
   t \rightarrow \frac{1}{\omega}t, \quad x &\rightarrow \delta\frac{1}{k} \xi, \quad z \rightarrow \frac{1}{k}z, \quad \phi \rightarrow \frac{\omega}{k^2}\phi, \quad (p + \rho g z) \rightarrow \frac{\rho g}{k} p \\
    u &\rightarrow \delta\frac{\omega}{k} u , \quad w \rightarrow \frac{\omega}{k} w, \quad  \theta \rightarrow \theta (\theta_w-\theta_m) + \theta_m, \quad 
\end{aligned}
    \label{eq:scaling_NS_ADE}
\end{equation}
where $\theta_m$ is the ice melting temperature and $\theta_w$ is the ambient water temperature. We have introduced a stretched coordinate $\xi$ for the wall-normal horizontal direction based on the boundary layer thickness, which is expected to be $O(\sqrt{\nu/\omega})$. Thus, $\delta = \sqrt{k^2\nu/\omega}$ is the dimensionless boundary layer thickness. We have also magnified the wall-normal velocity by a factor equal to the inverse boundary layer thickness. By using the boundary layer thickness in making the wall-normal horizontal coordinate and velocity dimensionless in this way, we ensure that the dimensionless horizontal dynamics are of the same order as the dimensionless vertical dynamics inside the boundary layer. The resulting dimensionless equations are given by
\begin{subequations}
    \begin{align}
        \text{continuity}&:\qquad \frac{\partial u}{\partial\xi} +  \frac{\partial w}{\partial z}=0, \\
        x\text{-momentum}&:\qquad \, \frac{\partial u}{\partial t} + u\frac{\partial u}{\partial\xi} + w\frac{\partial u}{\partial z}= - \delta^{-2} \frac{\partial p}{\partial\xi} + \frac{\partial^2 u}{\partial\xi^2} + \delta^{2} \, \frac{\partial^2 u}{\partial z^2}, \\
        z\text{-momentum}&:\qquad \frac{\partial w}{\partial t} +  u\frac{\partial w}{\partial\xi} + w\frac{\partial w}{\partial z}=- \frac{\partial p}{\partial z} + \frac{\partial^2w}{\partial\xi^2} + \delta^{2} \,\frac{\partial^2w}{\partial z^2}, \\
        \text{heat transport}&:\qquad \frac{\partial\theta}{\partial t} + u\frac{\partial\theta}{\partial \xi} + w\frac{\partial\theta}{\partial z}=\mathrm{Pr}^{-1}\,\frac{\partial^2\theta}{\partial\xi^2}+\mathrm{Pe}^{-1}\,\frac{\partial^2\theta}{\partial z^2},
    \label{eq:dimensionless_ADE}
    \end{align}
    \label{eq:dimensionless_NS_ADE1}%
\end{subequations}
where $\mathrm{Pr}={\nu}/{\alpha}$ is the Prandtl number, and $\mathrm{Pe}={\omega}/(\alpha{k^2})$ is the P\'{e}clet number, which together with the wave steepness $\varepsilon  = ka$ and the dimensionless boundary layer thickness $\delta$, are the dimensionless parameters in the problem so far.

The boundary conditions must account for the expected melt and changing position of the ice--water interface. We write the changing position of the ice--water interface as $\xi = m(z,t)$ with $m(z,t=0) = 0$ denoting an initial vertical ice--water interface at $\xi = 0$. The velocity boundary conditions are zero velocity at the ice--water interface and that the velocity tends toward the inviscid solution far outside the viscous boundary layer. The temperature boundary conditions are that it is equal to the melting temperature at the ice--water interface and tends toward the ambient water temperature far outside the thermal boundary layer. Additionally, the temperature field must also obey the Stefan boundary condition \citep[see \textit{e.g.},][]{AzizNa84, Mei09}, which states that the rate at which the ice--water interface melts is set by the balance between the flux of thermal energy into the ice and the latent heat required to melt the ice. All together, these boundary conditions are given in dimensionless form by
\begin{subequations}
    \begin{align}
        \xi = m &:\qquad (u, w, \theta) = (0, 0, 0), \label{eq:xzeroBC} \\
        \xi \rightarrow \infty &:\qquad (u, w, \theta) \rightarrow (0, i 2 \varepsilon e^{z}e^{it},1), \label{eq:xinftyBC} \\
        \xi = m &:\qquad \frac{dm}{dt} = - \mathrm{Ste} \, \mathrm{Pr}^{-1} \frac{\partial \theta}{\partial \xi}. \label{eq:StefanBC}
    \end{align}
    \label{eq:NS_ADE_BCs1}%
\end{subequations}
Here, $\mathrm{Ste} = c_p \lvert \theta_w - \theta_m \lvert /L$ is the Stefan number, with $c_p$ being the specific heat capacity of the water and $L$ being the latent heat of fusion of the ice.

To obtain the wave-induced ice melt rate, we must solve \eqref{eq:dimensionless_NS_ADE1} subject to boundary conditions (\ref{eq:xzeroBC}, \ref{eq:xinftyBC}) to obtain the temperature field and then calculate the melt rate using \eqref{eq:StefanBC}. 

By considering the dimensionless parameters in the problem, we note that the material properties of ice and water lead to a large Prandtl number ($\mathrm{Pr} \gtrsim 5$), a small Stefan number ($\mathrm{Ste} \lesssim 0.25$), and a small dimensionless boundary layer thickness ($\delta \lesssim 10^{-2}$). Also, small-amplitude wave theory and the onset of wave breaking restrict the wave steepness to be small ($\varepsilon \lesssim 0.25$). Based on these limits, we introduce the scalings
\begin{equation}
\varepsilon  \rightarrow \epsilon \varepsilon , \, \mathrm{Pr}^{-1}\rightarrow \epsilon^{2} \mathrm{Pr}^{-1}, \, \delta \rightarrow \epsilon \delta,  \, \mathrm{Pe}^{-1} = \delta^{2} \mathrm{Pr}^{-1} \rightarrow \epsilon^{4} \mathrm{Pe}^{-1}, \, \mathrm{Ste} \rightarrow \epsilon \mathrm{Ste},
\label{eq:scalings1}
\end{equation}
where $\epsilon$ is small and acts as an ordering parameter.

Next, we expand the solution variables as
\begin{subequations}
\begin{align}
x\text{-velocity}&:\quad u=u(\xi,z,t,T)= \epsilon u_1 +\epsilon^2 u_2 +\epsilon^3 u_3 + O(\epsilon^4), \\
z\text{-velocity}&:\quad w=w(\xi,z,t,T)= \epsilon w_1 +\epsilon^2 w_2 + \epsilon^3 w_3 + O(\epsilon^4), \\
\text{pressure}&:\quad p=p(\xi,z,t,T)= \epsilon p_1 +\epsilon^2 p_2 + \epsilon^3 p_3 + O(\epsilon^4), \\
\text{temperature}&:\quad \theta=\theta(\xi,z,t,T)=\theta_0 + \epsilon\theta_1 + \epsilon^2\theta_2 + \epsilon^3 \theta_3 + O(\epsilon^4), \\
\text{ice--water interface}&:\quad m = m(\xi,z,t,T) = \epsilon m_1 + \epsilon^2 m_2 + \epsilon^3 m_3 + O(\epsilon^4),
\end{align}
\label{eq:expansions1}%
\end{subequations}
where each variable is now a function of both fast time $t$ and slow time $T = \epsilon^2 t$, with the two times being separate independent variables and the time derivative becoming $\partial / \partial t \rightarrow \partial / \partial t  + \epsilon^2 \partial / \partial T $. Note, the expansion does not include $O(\epsilon^0)$ terms for the velocities, pressure, and the ice--water interface position. This is because using the scalings \eqref{eq:scalings1} in the inviscid solution shows that the wave-induced velocities and dynamic pressures are $O(\epsilon)$ or higher. The same is true for the ice--water position, as can be seen from using the scalings \eqref{eq:scalings1} and the multi-timescale expansion in the Stefan boundary condition \eqref{eq:StefanBC}. 

Using the scalings in \eqref{eq:scalings1} and ${\partial}/{\partial t}\rightarrow {\partial}/{\partial t}+\epsilon^2{\partial}/{\partial T}$, \eqref{eq:dimensionless_NS_ADE1} gives
\begin{subequations}
    \begin{align}
        \text{continuity}&:\quad \frac{\partial u}{\partial\xi} + \frac{\partial w}{\partial z}=0,\\
        x\text{-momentum}&:\quad \, \frac{\partial u}{\partial t} + \epsilon^{2} \frac{\partial u}{\partial T} +  u\frac{\partial u}{\partial\xi} + w\frac{\partial u}{\partial z} =  - \epsilon^{-2} \delta^{-2} \frac{\partial p}{\partial\xi} + \frac{\partial^2 u}{\partial\xi^2} + \epsilon^2 \delta^2 \frac{\partial^2 u}{\partial z^2}, \\
        z\text{-momentum}&:\quad \frac{\partial w}{\partial t} + \epsilon^{2} \frac{\partial w}{\partial T} + u\frac{\partial w}{\partial\xi} + w\frac{\partial w}{\partial z} = - \frac{\partial p}{\partial z} +  \frac{\partial^2w}{\partial\xi^2} + \epsilon^2 \delta^{2} \frac{\partial^2w}{\partial z^2}, \\
        \text{heat transport}&:\quad \frac{\partial\theta}{\partial t} + \epsilon^{2} \frac{\partial\theta}{\partial T} + u\frac{\partial\theta}{\partial \xi} +  w\frac{\partial\theta}{\partial z} = \epsilon^2 \mathrm{Pr}^{-1}\frac{\partial^2\theta}{\partial\xi^2} + \epsilon^4 \mathrm{Pe}^{-1}\frac{\partial^2\theta}{\partial z^2}.
    \label{eq:dimensionless_ADE}
    \end{align}
    \label{eq:dimensionless_NS_ADE_scaled}
\end{subequations}

Inserting \eqref{eq:expansions1} into \eqref{eq:dimensionless_NS_ADE_scaled} gives an ordered set of equations at different powers of $\epsilon$, which can be solved sequentially subject to boundary conditions \eqref{eq:NS_ADE_BCs1} as we now show. 

At $O(\epsilon^{-1})$, we find that only the $x$-momentum equation is relevant and it gives
\[\frac{\partial p_1}{\partial\xi} = 0.\]
This shows that the pressure in the boundary layer is uniform at this order and must be the same as the pressure outside the boundary layer. The pressure outside the boundary layer follows Bernoulli's equation, which when made dimensionless according to \eqref{eq:scaling_NS_ADE} and with the understanding that the inviscid horizontal velocity is zero, is given by
\begin{equation}
\frac{\partial \phi}{\partial t} + p_I + \tfrac{1}{4} w_I (w_I^*) = 0.
\label{eq:Bernoulli}
\end{equation}
In the last term, we have written $\tfrac{1}{2}[\Real(w_I)]^2$ using complex conjugates to ensure that only the real parts are retained. 
Substituting the pressure expansion into \eqref{eq:Bernoulli} using the scalings \eqref{eq:scalings1} gives $p_1 = -\left.{\partial \phi}/{\partial t}\right|_{x=0} =  2 \varepsilon e^{z} e^{it}$.

At $O(\epsilon^{0})$, we find
\begin{align*}
     x\text{-momentum}&:\qquad \frac{\partial p_2}{\partial\xi } =0  \\
    \text{Heat transport}&:\qquad \frac{\partial \theta_0}{\partial t} = 0
\end{align*}
From the $x$-momentum equation, we see that again the pressure in the boundary layer is uniform at this order and equal to the pressure outside the boundary layer. Substituting the pressure expansion into \eqref{eq:Bernoulli} using the scalings \eqref{eq:scalings1} a second time gives $p_2 = - \tfrac{1}{4} (w_I)(w_I^*)$. The heat transport equation shows that the leading order temperature solution is only a function of the spatial coordinates and slow time, \textit{i.e.}, $\theta_0 = \theta_0(\xi,z,T)$.

At $O(\epsilon^{1})$, we find
\begin{align*}
    \mathrm{Continuity}&:\qquad \frac{\partial u_1}{\partial\xi} + \frac{\partial w_1}{\partial z}=0\\
    x\text{-momentum}&:\qquad  \frac{\partial u_1}{\partial t} = - \delta^{-2}\frac{\partial p_3}{\partial\xi } + \frac{\partial^2 u_1}{\partial\xi^2}  \\
    z\text{-momentum}&:\qquad  \frac{\partial w_1}{\partial t} = -  2 \varepsilon e^{z} e^{it} +\frac{\partial^2 w_1}{\partial\xi^2} \\
    \text{Heat transport}&:\qquad \frac{\partial\theta_1}{\partial t} +  u_1 \frac{\partial\theta_0}{\partial \xi} +  w_1 \frac{\partial\theta_0}{\partial z} =0
\end{align*}
Starting with the $z$-momentum equation, we solve for $w_1$ using the substitution $\tilde{w}_1 = i2\varepsilon e^{z}e^{it} - w_1$, which then gives ${\partial \tilde{w}_1}/{\partial t}={\partial^2 \tilde{w}_1}/{\partial\xi^2}$ subject to boundary conditions $\tilde{w}_1 = i2\varepsilon e^{z}e^{it}$ at $\xi=m$ and $\tilde{w}_1 \rightarrow 0$ at $\xi \rightarrow \infty$. Writing $\tilde{w}_1=F_0(\xi)e^{it}$, the equation to solve becomes $iF_0(\xi)e^{it}=F''_0(\xi)e^{it}$, which has the solution
\[F_0(\xi)=Ae^{(1+i)\xi/\sqrt{2}}+Be^{-(1+i)\xi/\sqrt{2}}.\]
The boundary condition at $\xi\rightarrow\infty$ shows $A=0$. The boundary condition at $\xi=m$ can be simplified via $\tilde{w}_1 (\xi = m) = \tilde{w}_1 (\xi = 0) + m \left.{\partial \tilde{w}_1}/{\partial \xi}\right|_{\xi=0} + \text{h.o.t.}= \tilde{w}_1 (\xi = 0) + O(\epsilon^2) = i2\varepsilon e^{z}e^{it}$ to give $B=i2 \varepsilon e^z$. Therefore, the $O(\epsilon^{1})$ solution for the vertical velocity is given by the real part of
\begin{equation}
    w_1=i 2 \varepsilon e^z e^{it}\left(1 - e^{-(1+i)\xi/\sqrt{2}}\right).
    \label{eq:w1_soln}
\end{equation}

At this point, it is useful to note that the boundary conditions applied at the ice--water interface $\xi = m$ can be approximated using a Taylor series expansion to apply at the initial ice--water interface $\xi=0$ with an error that is one order higher in $\epsilon$ since the leading order scaling for the interface position is $m = O(\epsilon)$. Below, we use this approximation implicitly without further comment.

With $w_1$ found, we can solve for $u_1$ using the continuity equation via $u_1 = -\int_0^\xi \frac{\partial{w_1}}{{\partial z}}\, d\xi$, which gives
\begin{equation}
    u_1 = -i 2 \varepsilon e^z e^{it} \left[ \xi-\frac{\sqrt{2}}{1+i} \left( 1 - e^{-(1+i)\xi/\sqrt{2}} \right) \right]
    \label{eq:u1_soln}
\end{equation}
where again it is implied that the solution is given by the real part of the expression.

We could use the $x$-momentum equation to solve for $p_3$, but since it is not required, we move on the heat transport equation. The $O(\epsilon^1)$ heat transport equation shows that $\theta_1$ is given by the advection of $\theta_0$ by $(u_1,w_1)$ with no diffusion, so we can solve for $\theta_1$ since $u_1$ and $w_1$ are both known. The zero-mean solution is given by the time integrals
\[
\theta_1 = - \frac{\partial\theta_0}{\partial \xi}  \int u_1 \,dt - \frac{\partial\theta_0}{\partial z} \int w_1 \, dt,
\]
where we have used the fact that $\theta_0$ is independent of the fast time $t$. The solution is the real part of
\begin{equation}
    \theta_1 = i \left(u_1 \frac{\partial\theta_0}{\partial \xi} + w_1  \frac{\partial\theta_0}{\partial z} \right),
    \label{theta1_soln}
\end{equation}
which is in terms of the slow-time temperature solution $\theta_0$ that is yet to be found.

At $O(\epsilon^2)$, we find
\begin{align*}
    \mathrm{Continuity}&:\qquad \frac{\partial u_2}{\partial\xi} + \frac{\partial w_2}{\partial z}=0\\
    x\text{-momentum}&:\qquad  \frac{\partial u_2}{\partial t} +  u_1\frac{\partial u_1}{\partial\xi} +  w_1\frac{\partial u_1}{\partial z} =  - \delta^{-2} \frac{\partial p_4}{\partial\xi} + \frac{\partial^2 u_2}{\partial\xi^2}  \\
    z\text{-momentum}&:\qquad  \frac{\partial w_2}{\partial t} + u_1 \frac{\partial w_1}{\partial\xi} + w_1 \frac{\partial w_1}{\partial z} = w_I \frac{\partial w_I}{\partial z} + \frac{\partial^2 w_2}{\partial\xi^2} \\
    \text{Heat transport}&:\qquad \frac{\partial\theta_2}{\partial t} +  \frac{\partial\theta_0}{\partial T} + u_2 \frac{\partial\theta_0}{\partial \xi} + u_1 \frac{\partial\theta_1}{\partial \xi} + w_2 \frac{\partial\theta_0}{\partial z} + w_1 \frac{\partial\theta_1}{\partial z} =  \mathrm{Pr}^{-1}\frac{\partial^2\theta_0}{\partial\xi^2}
\end{align*}

Wave-averaging the $z$-momentum equation provides a governing equation for the steady streaming component of the vertical velocity
\[
\frac{\partial^2 \overline{w_2}}{\partial \xi^2} = \overline{\left({u_1\frac{\partial w_1}{\partial\xi}} + {w_1 \frac{\partial w_1}{\partial z}} - w_I \frac{\partial w_I}{\partial z} \right)},
\]
where the wave-averaging operator is defined as $\overline{f} = (1/2\pi)\int_0^{2\pi} f \, \mathrm{d}t$. Using complex conjugates to ensure that only the real parts of the complex expressions are retained and then only the terms that lead to the steady components gives
\[
\frac{\partial^2 \overline{w_2}}{\partial\xi^2} = \tfrac{1}{2} \left[u_1\frac{\partial w_1^*}{\partial\xi} + w_1 \frac{\partial w_1^*}{\partial z} - w_I \frac{\partial w_I^*}{\partial z} \right].
\]
Integrating twice with the boundary conditions that $\overline{w_2} = 0$ at $\xi=0$ and $\partial \overline{w_2}/\partial \xi = 0$ at $\xi \rightarrow \infty$ brings us to
\begin{multline}
    \overline{w_2} = \varepsilon^2 e^{2z} \bigg[-3 + 3i + (1+i)e^{-\sqrt{2}\xi}  + 2(1-3i)e^{-(1-i)\xi/\sqrt{2}}  \\  + 2i e^{-(  1+i)\xi/\sqrt{2}} - \sqrt{2}(1+i)\xi e^{-(1-i)\xi/\sqrt{2}} \bigg],
\end{multline}
where again the solution is given by the real part.

Wave-averaging the continuity equation produces an expression for the steady streaming component of the horizontal velocity: $\overline{u_2}= -\int_0^\xi \frac{\partial\overline{w_2}}{\partial z}\,d\xi$, with the solution given by the real part of
\begin{multline}
    \overline{u_2} = -2\varepsilon^2 e^{2z} \bigg[(-3 + 3i)\xi + \frac{\sqrt{2}}{2}(1+i)(1 - e^{-\sqrt{2}\xi}) \\  + \sqrt{2}(5-3i)(1 - e^{-(1-i)\xi/\sqrt{2}}) + \sqrt{2}(1+i)(1 - e^{-(  1+i)\xi/\sqrt{2}}) + 2i \xi e^{-(1-i)\xi/\sqrt{2}} \bigg].
\end{multline}

Wave-averaging the heat transport equation gives
\[
\frac{\partial\theta_0}{\partial T} + \overline{u_2} \frac{\partial\theta_0}{\partial \xi} + \overline{w_2} \frac{\partial\theta_0}{\partial z} + \overline{u_1 \frac{\partial\theta_1}{\partial \xi}} + \overline{w_1 \frac{\partial\theta_1}{\partial z}} = \mathrm{Pr}^{-1}\,\frac{\partial^2\theta_0}{\partial\xi^2}
\]
Again, using complex conjugates to ensure that only the real parts of the complex expressions are retained and then only the terms that lead to the steady components results in
\begin{multline}
\frac{\partial\theta_0}{\partial T} + \left[ \overline{u_2} -i\tfrac{1}{2} \left(w_1 {\frac{\partial u_1}{\partial z}}^* + u_1 {\frac{\partial u_1}{\partial \xi}}^*\right) \right]\frac{\partial\theta_0}{\partial \xi} + \left[ \overline{w_2} -i\tfrac{1}{2} \left(w_1 {\frac{\partial w_1}{\partial z}}^* + u_1 {\frac{\partial w_1}{\partial \xi}}^*\right) \right]\frac{\partial\theta_0}{\partial z} \\ = \mathrm{Pr}^{-1}\,\frac{\partial^2\theta_0}{\partial\xi^2}
\end{multline}
We see that the wave-averaged temperature is governed by temporal evolution with respect to slow time, advection by the terms in square brackets, and diffusion in the wall-normal direction. We also see that the advection terms are comprised of not only the wave-averaged velocities (first term in each square bracket), but also additional contributions (terms inside the round brackets in each square bracket) that correspond to the Stokes drift that arises when computing Lagrangian transport of passive scalars in a wavy environment \citep{Buhler09}.

For our purposes, we can rewrite the wave-averaged heat transport equation as
\begin{equation}
   \frac{\partial\Theta}{\partial T} + U\frac{\partial\Theta}{\partial \xi} + W\frac{\partial\Theta}{\partial z} = \mathrm{Pr}^{-1}\frac{\partial^2\Theta}{\partial\xi^2},
   \label{eq:WA-ADE1}
\end{equation}
where the capital variables denote wave-averaged quantities: $\Theta = \theta_0$ is the wave-averaged temperature, $T$ is the wave-averaged time as before, and $(U, W)$ are the effective wave-averaged advection velocities given by the real parts of
\begin{subequations}
\begin{align}
U &=  \overline{u_2} -i\tfrac{1}{2} \left(w_1 {\frac{\partial u_1}{\partial z}}^* + u_1 {\frac{\partial u_1}{\partial \xi}}^*\right) \\
W &=  \overline{w_2} -i\tfrac{1}{2} \left(w_1 {\frac{\partial w_1}{\partial z}}^* + u_1 {\frac{\partial w_1}{\partial \xi}}^*\right).
\end{align}
\label{eq:WA-advectionvel}
\end{subequations}
The Stokes drift terms (in brackets) can be computed from solutions of $u_1$ and $w_1$ to give
\begin{subequations}
\begin{align}
-i\tfrac{1}{2} \left(w_1 {\frac{\partial u_1}{\partial z}}^* + u_1 {\frac{\partial u_1}{\partial \xi}}^*\right)  &= -i 2  \varepsilon^2 e^{2z} \bigg[ \sqrt{2} - 2\xi + \sqrt{2}e^{-\sqrt{2}\xi}  \nonumber \\
& \qquad - e^{-(1-i)\xi/\sqrt{2}} \left( \sqrt{2} - \xi \right) - e^{-(1+i)\xi/\sqrt{2}} \left( \sqrt{2} - \xi \right) \bigg], \\
-i\tfrac{1}{2} \left(w_1 {\frac{\partial w_1}{\partial z}}^* + u_1 {\frac{\partial w_1}{\partial \xi}}^*\right) &= -i 2 \varepsilon^2 e^{2z} \bigg[ 1 + e^{-\sqrt{2}\xi} (1+i) - e^{-(1+i)\xi/\sqrt{2}} \nonumber \\ 
 &  \qquad \qquad \qquad - e^{-(1-i)\xi/\sqrt{2}} \left( 1 + i + \xi \frac{(1-i)}{\sqrt{2}} \right) \bigg].
\end{align}
\end{subequations}

Now, \eqref{eq:WA-ADE1} with $U, W$ given by \eqref{eq:WA-advectionvel} is the final advection-diffusion equation for the wave-averaged temperature in the boundary layer. We pause here to comment on three aspects of this result. First, we note that \eqref{eq:WA-ADE1} is applicable only near the surface where $-z = O(1)$ or smaller. For large depths  ($z \rightarrow -\infty$), the wave-induced velocities vanish ($U, W \rightarrow 0$) and \eqref{eq:WA-ADE1} reduces to a one-dimensional diffusion equation. 

Second, our solutions for $U, W$ imply that $U(\xi \rightarrow \infty) \rightarrow \infty$, which is nonphysical, and that $W(\xi \rightarrow \infty) = -3\varepsilon^2 e^{2z}$, which does not match the inviscid region outside the boundary layer. In reality, the wave-averaged boundary layer flow would eventually lead to a modification of the inviscid and irrotational flow outside the boundary layer, as shown by \citet{LH53}, and all velocities would remain bounded and match to this modified inviscid and irrotational flow. The modified flow will satisfy different field equations for its streamfunction based on the ratio of the wave amplitude to the boundary layer thickness. Most applications, including ours, are likely to fall into the so-called `convection' regime where this ratio is large. However, as \citet{LH53} makes clear, the boundary layer streaming itself is only dependent on the first-order oscillatory motion and the local boundary conditions, consistent with how we have calculated it.

Last, we have neglected the full complexity of the free-surface boundary conditions. We do not satisfy the second-order free surface boundary conditions and we ignore viscous effects, including the horizontal streaming, that would occur there. We have also neglected surface tension, which would modify the free-surface boundary conditions at the moving contact line at the ice--water interface. Theoretical treatments of the flow dynamics near the moving contact line, subject to both viscous and surface tension effects, not to mention boundary layer streaming and separation, quickly become complicated \citep[see][]{MeiLiu73, Hocking87, Miles90}. Additionally, laboratory measurements of the contact line dynamics and adjacent flow for wave reflection at a vertical wall by \citet{Park12} show the existence of a downward jet flow during the descending phase of the free surface. None of these effects are included here, but the downward jets may compensate for the effects we have ignored by producing a downward flux of heat similar to our vertical mass transport velocity.

Putting these complexities aside, we now proceed with the problem of computing the wave-induced melt rate by turning to the Stefan condition. Inserting the expansion for the interface position $m$ into the Stefan condition shows that the first melting occurs at $O(\epsilon^3)$ with a wave-averaged melting given by $d m_1/dT$. Writing $M = m_1$ for the wave-averaged interface position, the wave-averaged Stefan condition becomes
\begin{equation}
\frac{dM}{dT} = - \mathrm{Ste} \, \mathrm{Pr}^{-1} \frac{\partial \Theta}{\partial \xi} \bigg\lvert_{\xi=m} = - \mathrm{Ste} \, \mathrm{Pr}^{-1} \frac{\partial \Theta}{\partial \xi} \bigg\lvert_{\xi=0} + \, \text{h.o.t.}
\label{eq:StefanBC-WA1}
\end{equation}
Using this, the unsteady term in \eqref{eq:WA-ADE1} can be written as
\[
 \frac{\partial\Theta}{\partial T} =  \frac{\partial\Theta}{\partial M}  \frac{d M}{d T} =  \frac{\partial\Theta}{\partial M} \left(- \mathrm{Ste} \, \mathrm{Pr}^{-1} \frac{\partial \Theta}{\partial \xi} \bigg\lvert_{\xi=0} \right) = - \mathrm{Ste} \, \mathrm{Pr}^{-1}  \frac{\partial\Theta}{\partial M} \frac{\partial \Theta}{\partial \xi} \bigg\lvert_{\xi=0}
\]
This turns the wave-averaged heat transport equation \eqref{eq:WA-ADE1} into
\begin{equation}
  -{\mathrm{Ste}}\,{\mathrm{Pr}^{-1}}  \frac{\partial\Theta}{\partial M} \frac{\partial \Theta}{\partial \xi} \bigg\lvert_{\xi=0} + U \frac{\partial\Theta}{\partial \xi} + W \frac{\partial\Theta}{\partial z} = \mathrm{Pr}^{-1}\frac{\partial^2\Theta}{\partial \xi^2}
  \label{eq:WA-ADE-R0}
\end{equation}
From the scalings in \eqref{eq:scalings1} and recalling that $(U, W) = O(\epsilon^2)$, we can see that the leading order dynamics for the wave-averaged temperature field are a steady balance between advection and horizontal diffusion given by
\begin{equation}
U \frac{\partial\Theta}{\partial \xi} + W \frac{\partial\Theta}{\partial z} = \mathrm{Pr}^{-1}\frac{\partial^2\Theta}{\partial \xi^2}.
\label{eq:WA-ADE-R1}
\end{equation}

We can find the melt rate by solving \eqref{eq:WA-ADE-R1} for $\Theta$ and inserting this solution into \eqref{eq:StefanBC-WA1}. In the main text, we solve a simplified, approximate version of \eqref{eq:WA-ADE-R1} that neglects the horizontal advection and considers the vertical advection only due to the Eulerian streaming velocity at the edge of the boundary layer.

\bibliographystyle{jfm}
\bibliography{waveicemelt.bib}

\end{document}